\documentclass[%
 reprint,
 amsmath,amssymb,
 aps,
twocolumn,prc,nofootinbib,floatfix,superscriptaddress
]{revtex4-2}

\usepackage{graphicx}% Include figure files
\usepackage{dcolumn}% Align table columns on decimal point
\usepackage{bm}% bold math
\begin{document}

\preprint{APS/123-QED}

\title{First Penning trap mass measurement of $^{36}$Ca}

\author{J. Surbrook}
\affiliation{%
 Department of Physics and Astronomy, Michigan State University, East Lansing, Michigan 48824, USA
}%
\affiliation{
 National Superconducting Cyclotron Laboratory, East Lansing, Michigan 48824, USA
}

\author{G. Bollen}
\affiliation{%
 Department of Physics and Astronomy, Michigan State University, East Lansing, Michigan 48824, USA
}%
\affiliation{
 Facility for Rare Isotope Beams, East Lansing, Michigan 48824, USA
}%

\author{M. Brodeur}
\affiliation{
 Department of Physics, University of Notre Dame, Notre Dame, Indiana 46556, USA
}%

\author{A. Hamaker}
\affiliation{%
 Department of Physics and Astronomy, Michigan State University, East Lansing, Michigan 48824, USA
}%
\affiliation{
 National Superconducting Cyclotron Laboratory, East Lansing, Michigan 48824, USA
}

%\author{E. Leistenschneider}
%\affiliation{
% National Superconducting Cyclotron Laboratory, East Lansing, Michigan 48824, USA
%}%

\author{D. P{\'e}rez-Loureiro}
\affiliation{
 National Superconducting Cyclotron Laboratory, East Lansing, Michigan 48824, USA
}%

\author{D. Puentes}
\affiliation{%
 Department of Physics and Astronomy, Michigan State University, East Lansing, Michigan 48824, USA
}%
\affiliation{
 National Superconducting Cyclotron Laboratory, East Lansing, Michigan 48824, USA
}

\author{C. Nicoloff}
\affiliation{%
 Department of Physics and Astronomy, Michigan State University, East Lansing, Michigan 48824, USA
}%
\affiliation{
 National Superconducting Cyclotron Laboratory, East Lansing, Michigan 48824, USA
}

\author{M. Redshaw}
\affiliation{
 Department of Physics, Central Michigan University, Mount Pleasant, Michigan 48859, USA
}%
\affiliation{
 National Superconducting Cyclotron Laboratory, East Lansing, Michigan 48824, USA
}%

\author{R. Ringle}
\affiliation{
 National Superconducting Cyclotron Laboratory, East Lansing, Michigan 48824, USA
}%

\author{S. Schwarz}
\affiliation{
 National Superconducting Cyclotron Laboratory, East Lansing, Michigan 48824, USA
}%

\author{C.S. Sumithrarachchi}
\affiliation{
 National Superconducting Cyclotron Laboratory, East Lansing, Michigan 48824, USA
}%

\author{L.J. Sun}
\affiliation{
 National Superconducting Cyclotron Laboratory, East Lansing, Michigan 48824, USA
}%

\author{A.A. Valverde}
\affiliation{
 Department of Physics and Astronomy, University of Manitoba, Winnipeg, Manitoba R3T 2N2, Canada
}%

\author{A.C.C. Villari}
\affiliation{
 Facility for Rare Isotope Beams, East Lansing, Michigan 48824, USA
}%

\author{C. Wrede}
\affiliation{%
 Department of Physics and Astronomy, Michigan State University, East Lansing, Michigan 48824, USA
}%
\affiliation{
 National Superconducting Cyclotron Laboratory, East Lansing, Michigan 48824, USA
}

\author{I.T. Yandow}
\affiliation{%
 Department of Physics and Astronomy, Michigan State University, East Lansing, Michigan 48824, USA
}%
\affiliation{
 National Superconducting Cyclotron Laboratory, East Lansing, Michigan 48824, USA
}

\date{\today}% It is always \today, today,

\begin{abstract}
\begin{description}
\item[Background]
Isobaric quintets provide the best test of the isobaric multiplet mass equation (IMME) and can uniquely identify higher order corrections suggestive of isospin symmetry breaking effects in the nuclear Hamiltonian. The Generalized IMME (GIMME) is a novel microscopic interaction theory that predicts an extension to the quadratic form of the IMME. Only the $A=20, 32$ $T=2$ quintets have the exotic $T_z = -2$ member ground state mass determined to high-precision by Penning trap mass spectrometry. 
\item[Purpose]
To establish $A=36$ as the third high-precision $T=2$ isobaric quintet with the $T_z = -2$ member ground state mass measured by Penning trap mass spectrometry and provide the first test of the predictive power of the GIMME. 
\item[Method]
A radioactive beam of neutron-deficient $^{36}$Ca was produced by projectile fragmentation at the National Superconducting Cyclotron Laboratory. The beam was thermalized and the mass of $^{36}$Ca$^+$ and $^{36}$Ca$^{2+}$ measured by the Time of Flight - Ion Cyclotron Resonance method in the LEBIT 9.4 T Penning trap.
\item[Results]
We measure the mass excess of $^{36}$Ca to be ME$ = -6483.6(56)$ keV, an improvement in precision by a factor of 6 over the literature value. The new datum is considered together with evaluated nuclear data on the $A=36$, $T=2$ quintet. We find agreement with the quadratic form of the IMME given by isospin symmetry, but only coarse qualitative agreement with predictions of the GIMME.
\item[Conclusion]
A total of three isobaric quintets have their most exotic members measured by Penning trap mass spectrometry. The GIMME predictions in the $T = 2$ quintet appear to break down for $A = 32$ and greater.  

\end{description}
\end{abstract}

\maketitle

%\tableofcontents

\section{\label{sec:intro}Introduction}

The concept of isospin symmetry, put forth by Heisenberg, is a generalization of the similarities of the proton and neutron under the influence of the strong nuclear force \cite{Heis32}. It treats the proton and neutron as degenerate states of the same hadronic particle and frames an elegant explanation to the similarity of \textit{p} and \textit{n} masses, \textit{np} and \textit{pp} nuclear scattering, and properties of atomic nuclei with the same number of total nucleons, $A$. Isospin introduces new quantum numbers \textbf{$T$} and \textbf{$T_z$}. By convention, $T_z = 1/2$ for the free neutron, $T_z=-1/2$ for the free proton. In atomic nuclei with $N$ neutrons and $Z$ protons, isospin coupling yields $T_z = (N-Z)/2$ and allows $T = |T_z|, |T_z| +1, ..., A/2$. Across isobars, states with similar properties can belong to isospin-degenerate multiplets and are called \textit{isobaric analog states} (IAS).

Under isospin symmetry, all members of a multiplet would have the same mass. This symmetry is broken by the mass difference between protons and neutrons and Coulomb interactions between nucleons. First-order perturbation theory gives a correction to the nuclear mass excess, the \textit{Isobaric Multiplet Mass Equation} (IMME) \cite{Wein59}:
\begin{equation*}
    \text{ME}(A,T,T_z) = a(A,T) + b(A,T) T_z + c(A,T) T_z^2
\end{equation*}
Coefficients are determined by either theoretical prediction or fitting to measured nuclear masses. Second-order Coulomb effects, 3-body interactions, and isospin mixing naturally extend the IMME by adding $d T_z^3$ and $e T_z^4$ terms, but these coefficients have generally been expected to be small ($\lesssim 1$ keV) \cite{Henl69,Jane69,Bert70}. Therefore, a need for large $d$ or $e$ coefficients to describe a multiplet suggests a breakdown of isospin symmetry. Recent work done with \textit{ab initio} methods (AV18 and AV14 interactions \cite{Wiri95} solved with the Brueckner theory \cite{Brue58}) introduce the \textit{Generalized IMME} (GIMME) that predicts $d$ coefficients of a few keV (and smaller $e$ coefficients) that vary over atomic number $A$. These non-zero coefficients arise from the density dependence of isospin non-conserving effects in the nuclear medium and Coulomb polarization effects in the common proton-neutron ``core" of a multiplet \cite{Dong18,Dong19}. 

Higher-order coefficients in the IMME are best searched for in $T=2$ quintets, where $d$ and $e$ coefficients can be uniquely determined simultaneously. Often, the largest challenge in measurement of a quintet's IMME coefficients is the difficulty of precision measurement of the neutron-deficient $T_z = -2$ member. The short half-lives and challenging production, due to proximity to the proton drip-line, of these members limit the range of experimental measurement techniques capable of keV-level precision. As such, only two quintets exist ($A=20, 32$) \cite{MacC14,AME2016,Glas15} with all members' ground state masses measured to keV-level precision. In both cases, $^{20}$Mg ($A=20,T_z=-2$) and $^{32}$Ar ($A=32,T_z=-2$) were measured by Penning trap mass spectrometry, the most precise method of determining atomic mass \cite{Blau06,Lunn03,AME2016}. The current mass excess of $^{36}$Ca, ME$ = -6450(40)$ keV, has remained unchallenged since the measurement of the $^{40}$Ca($^4$He,$^8$He)$^{36}$Ca neutron-knockout reaction \textit{Q}-value was published in 1977 \cite{Trib77,AME2016}. Another reaction measurement around 1998 remains unpublished \cite{Komi98,Gree99,Cagg99}. A Penning trap mass measurement of $^{36}$Ca would mark the third (and heaviest) IMME quintet with the challenging $T_z = -2$ member measured to high precision by Penning trap mass spectrometry. The complementary $A=36$ members are already known to a satisfactory precision through a combination of mass measurements involving Penning traps and careful particle threshold measurements \cite{AME2016}. Additionally, the GIMME predicts the $A=36$ quintet $d$ coefficient to have the largest departure from the quadratic IMME, making the mass measurement of $^{36}$Ca particularly relevant \cite{Dong19}.

\section{\label{sec:experiment}Experiment}
\begin{figure}[b]
\includegraphics[width=0.475\textwidth]{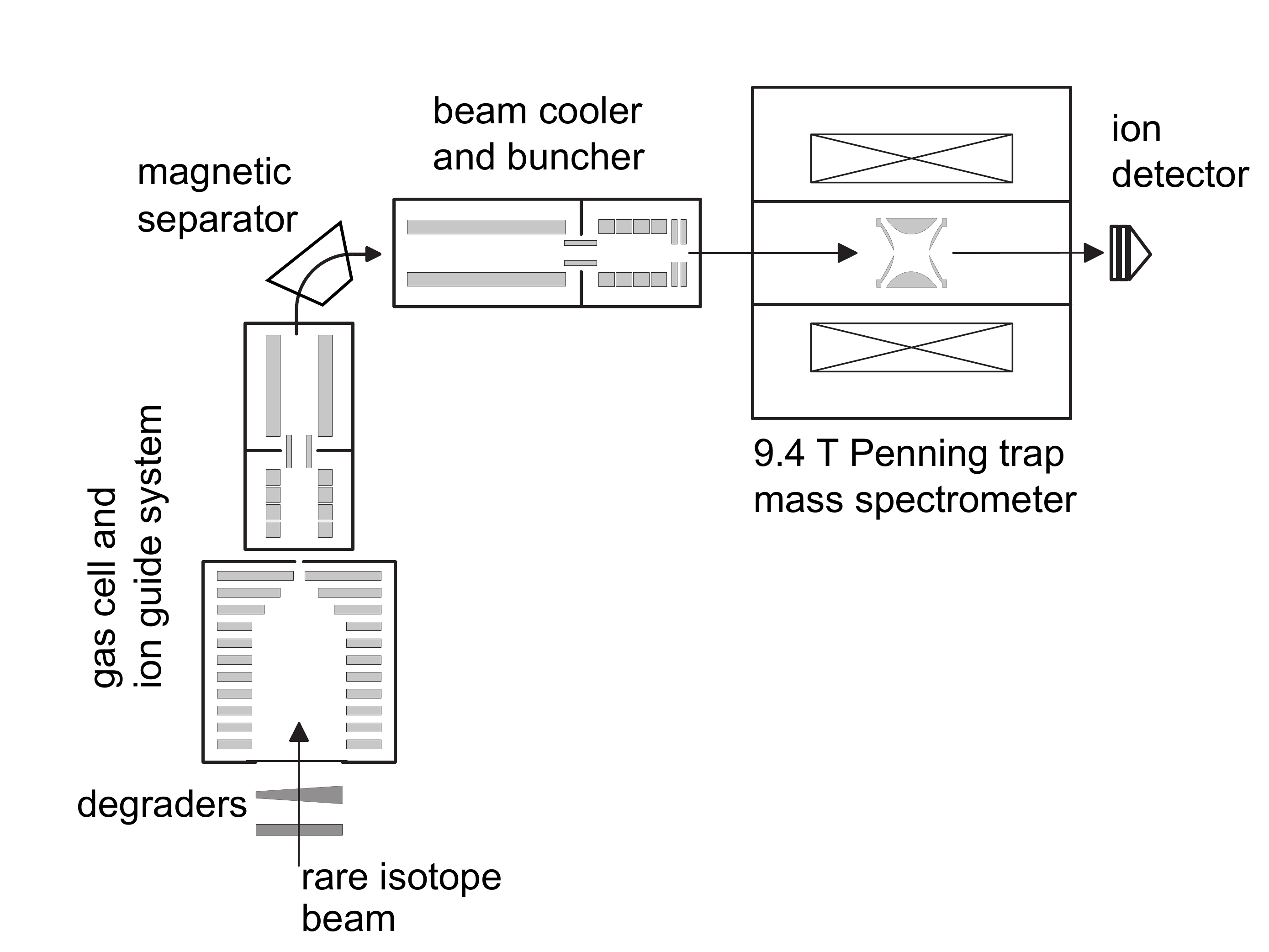}
\caption{\label{fig:lebitSchematic} Schematic of the NSCL gas cell and LEBIT facility. For further discussion, see Ref. \cite{Ring13}}
\end{figure}

At the National Superconducting Cyclotron Laboratory, a primary beam of $^{40}$Ca was accelerated to 140 MeV/nucleon with the Coupled Cyclotron Facility and impinged on a 658 mg/cm$^2$ thick target of natural beryllium to produce $^{36}$Ca. The mixed beam was transported to the A1900 \cite{Morr03}, where magnetic isotopic-selection transported a purified ($\sim 1\%$) beam of $^{36}$Ca to the NSCL gas cell \cite{Sumi20}. Aluminum degraders and high-purity helium gas in the gas cell (95 mbar) stopped the fast beam. Internal radio frequency quadrupole electrodes guided ions out of the gas cell volume to vacuum, where a dipole magnet of resolving power $\sim 1500$ selected ions of a specific mass-to-charge ratio $A/Q$. Ions were transported to the Low-Energy Beam and Ion Trap (LEBIT) facility, which is unparalleled among Penning trap facilities for its ability to measure the mass of short-lived rare isotopes produced by projectile fragmentation \cite{Ring13}. Fig. \ref{fig:lebitSchematic} shows a schematic of the gas cell and LEBIT facility.

In LEBIT, the mass of $^{36}$Ca was measured by the Time of Flight - Ion Cyclotron Resonance method. Following production, the ion beam is stopped, cooled, and accumulated to form ion bunches in low-pressure helium gas ($3\cdot 10^{-2}$ mbar cooling, $5\cdot 10^{-4}$ mbar bunching) \cite{Schw16}. After an accumulation time of 100-200 ms, a short $<10$ $\mu$s bunch is ejected. A fast electrostatic kicker provides additional $A/Q$ selection by time-of-flight (TOF) filtering before the ion bunch passes into the LEBIT Penning trap, a high-precision hyperbolic electrode trap system housed in a 9.4 T superconducting solenoidal magnet \cite{Ring07}. In the trap, the ion's motion is described by three eigenmotions with characteristic frequencies ($\nu_i$): trap-axial ($\nu_z$), magnetron ($\nu_-$), and modified-cyclotron ($\nu_+$), as described in Ref. \cite{Brow82}. The ion bunch is purified by applying a sum of dipole radio frequency electric field (RF) at the modified-cyclotron frequencies of identified contaminant ions and a broadband dipole RF with a 10 kHz window outside of the ion of interest's $\nu_+$  \cite{Reds13,Kwia15}. This dipole RF excites contaminant ions' modified-cyclotron motion out of the trap center, reducing the presence of trapped contaminant ions. Quadrupolar RF with a frequency $\nu_{RF}$ is applied in the trap near the ion of interest's cyclotron frequency $\nu_c = \nu_- + \nu_+$ to convert initial magnetron motion created by Lorentz steerers \cite{Ring07} to modified-cyclotron motion. Ions are ejected to a multi-channel plate detector outside of the 9.4 T magnet, where the ions' time-of-flight is measured. This procedure is repeated while scanning $\nu_{RF}$ near $\nu_c$ to create a TOF curve as in Fig. \ref{fig:TOF}. The ion's TOF is minimized for $\nu_{RF} = \nu_c$.

The measurement was performed across two settings. We first chose the gas cell dipole magnet to select $A/Q = 36$ and measured the cyclotron frequency of $^{36}$Ca$^+$ in the Penning trap. The reference measurement for this setting was $^{12}$C$_{3} ^+$ (Setting I). We then set the gas cell dipole magnet for $A/Q = 45$, selecting for molecular $^{36}$Ca(H$_2$O)$_3^{2+}$. A potential difference of 100 V was applied between the gas cell and the LEBIT cooler and buncher to liberate $^{36}$Ca$^{2+}$ via collision-induced disassociation \cite{Schu06}. The cyclotron frequency of $^{36}$Ca$^{2+}$ was measured in the Penning trap, using H$_2$O$^+$ as the reference (Setting II). Both settings excited the $^{36}$Ca ions for 50 ms per scan. Frequency measurements of the ion of interest and the reference ion were alternated in time, with no measurement running longer than 90 minutes.

\section{\label{sec:analysis}Analysis}

%\subsection{\label{subsec:massdetermination}Mass Analysis}
%The cyclotron frequency of an ion in a magnetic field, $B$, is related to its mass by 
%\begin{equation*}
%    \nu_c = \frac{qB}{m_{ion}}
%\end{equation*}{}

\begin{figure}[b]
\includegraphics[width=0.475\textwidth]{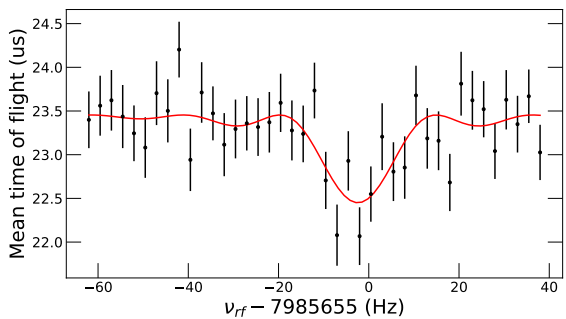}
\caption{\label{fig:TOF} Time-of-flight resonance of $^{36}$Ca$^{2+}$ with 50 ms excitation time. The smooth curve shows the fit of the theoretical curve, used to extract $\nu_c$ \cite{Koni95}.}
\end{figure}

\begin{figure}[b]
\includegraphics[width=0.475\textwidth]{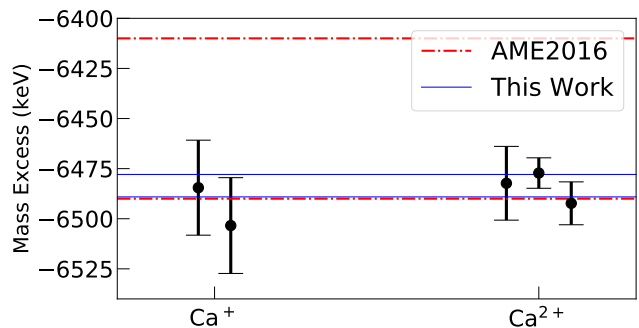}
\caption{\label{fig:MassExcesses} Mass excess of $^{36}$Ca determined by measurements of $^{36}$Ca$^+$ and $^{36}$Ca$^{2+}$ and their 1$\sigma$ error bars. The horizontal lines show the 1$\sigma$ error band from the AME2016 (dot-dashed red) \cite{AME2016,Trib77} and this work (solid blue). Note the near overlap of lower-bound lines near ME $\sim -6490$ keV.}
\end{figure}

We determine the mass of the ion of interest by calculating the ratio of the ion-of-interest (ion) cyclotron frequency and the interpolated cyclotron frequency of the reference ion (ref), $R = \nu_{\text{ref}}/\nu_{\text{ion}}$. Fig. \ref{fig:TOF} shows one time-of-flight spectrum used to extract a cyclotron frequency of $^{36}$Ca$^{2+}$. The analytic form of the TOF response curve of an ion excited in the Penning trap is presented in Ref. \cite{Koni95}. The full shape of the TOF curve was fit to data and the cyclotron frequency was extracted. The mass of the ion-of-interest is then $M = (q/q_{\text{ref}})\cdot R \cdot (m_{\text{ref}} - q_{\text{ref}}\cdot m_e) + q\cdot m_e$. For setting I, $q$ = 1 and for setting II, $q$ = 2. The reference ion was singly charged in both settings, so $q_{\text{ref}}$ = 1. We measured $R=1.00019369(50)$ for the Setting I and $R=1.00078041(18)$ for Setting II. Mass dependent systematic shifts, such as those due to inhomogeneities in the magnetic field or imperfections in the trap geometry, have been investigated at LEBIT and characterized at $\Delta R \sim 2 \cdot 10^{-10}$/u \cite{Guly15}. The use of $A/Q$ = 36 and 18 mass doublets, for setting I and II respectively, renders such mass-dependent systematic negligible. Effects due to special relativity are estimated to $\Delta R\leq 10^{-11}$. Non-linear fluctuations in the magnetic field have been shown to yield $\Delta R < 10^{-9}$/hour \cite{Ring07PRC}. Shifts due to ion-ion interactions are below our statistical uncertainty - more than 75\% of non-empty bunches contained a single ion and bunches with 5 or more ions were discarded from the analysis. As these systematic effects are well controlled relative to our statistical precision ($\sim 10^{-7}$), they were considered to be negligible in the analysis. While each setting has a limited number of measurements, the near-unity Birge ratio \cite{Birg32} (setting I - 0.56(33), setting II - 0.81(27) ) of each setting suggests that the assigned statistical error bars are appropriate. We used the most recent Atomic Mass Evaluation \cite{AME2016} for the mass of the reference ion in each setting to calculate the mass excess of $^{36}$Ca. The chemical binding energy ($\sim$ 10 eV) was treated as negligible. The two settings yield $ME_{\text{setting-I}} = -6494(17)$ keV and $ME_{\text{setting-II}} = -6482.3(59)$ keV, for an error-weighted averaged value of ME$ = -6483.6(56)$ keV, shown in Fig. \ref{fig:MassExcesses}. The agreement between the two $A/Q$ settings demonstrates the strong constraints on mass-dependent systematic effects.

%\begin{table}[b]%The best place to locate the table environment is directly after its first reference in text
%\caption{\label{tab:measurementValues} Mass excess of $^{36}$Ca determined by measurement of $^{36}$Ca$^+$ and $^{36}$Ca$^{2+}$ in this work.}
%\begin{ruledtabular}
%\begin{tabular}{cc}
%    Measurement                         & Mass Excess (keV) \\
%    \colrule
%    $^{36}$Ca$^+$       & $-6493.9(168)$ \\
%    $^{36}$Ca$^{2+}$    & $-6482.3(59)$ \\
%    Average           & $-6484(6)$ \\
%\end{tabular}
%\end{ruledtabular}
%\end{table}

    %\begin{table}[b]%The best place to locate the table environment is directly after its first reference in text
    %\caption{\label{tab:systematicEffects} Estimated systematic effects to our cyclotron frequency measurements.}
    %\begin{ruledtabular}
    %\begin{tabular}{cccc}
    %Description                         & $\Delta \nu_c/\nu_c$ \\
    %\colrule
    %non-linear \textbf{B}-field drift   & $< 5\cdot10^{-8}$/hour \\
    %space-charge interactions           & $< 10^{-8}$ \\
    %chemical binding energy             & $\sim 10^{-9}$ \\
    %\end{tabular}
    %\end{ruledtabular}
    %\end{table}

%\subsection{\label{subsec:IMMEanalysis}Update to the IMME}
\section{\label{sec:discussion}Discussion}
%\subsection{\label{subsec:IMME}IMME}
A literature search for updates to the $A=36$ quintet found the most recent IMME \cite{MacC14} and atomic mass \cite{AME2016} evaluations to be current. We adopted these values, replacing the previous $^{36}$Ca measurement by our new value with 6 times more precision, and performed quadratic, cubic, quadratic$+e T_z^4$, and quartic fits by $\chi^2$-minimization to extract coefficients of the extended IMME. Table \ref{tab:IMMEinput} gives the inputs for this analysis and Table \ref{tab:IMMEcoefs} presents the output coefficients. We find that the quadratic model describes the $A=36, T=2$ quintet well, with a chi-square per degree of freedom of $\chi^2/\nu = 0.65/2$ (p-value of 0.72). Furthermore, the quartic fit yields $d = -0.4(6)$ keV and $e = 0.2(6)$ keV, both of which are consistent with the IMME prediction of zero. $A=36$ now has the second-most precise measurement of $T=2$ IMME coefficients, following $A = 32$.

    \begin{table}[tp!]%The best place to locate the table environment is directly after its first reference in text
    \caption{\label{tab:IMMEinput} Mass excesses (ME) and excitation energy of the isobaric analog states (E$_{IAS}$) used to tabulate the mass excess of IAS states (ME$_\text{IAS}$) in our analysis. The $^{36}$Ca value is from the present measurement. Previous values are incorporated from Ref. \cite{MacC14,AME2016}.}
    \begin{ruledtabular}
    \begin{tabular}{cccc}
    Species & ME (keV)  &  E$_\text{IAS}$ (keV)  &  ME$_\text{IAS}$ (keV) \\
    \colrule
    $^{36}$Ca & $-6483.6(56)$     & --              & $-6483.6(56)$ \\
    $^{36}$K  & $-17417.1(3)$     & $4282.4(24)$    & $-13134.6(24)$ \\
    $^{36}$Ar & $-30231.540(27)$  & $10852.1(12)$   & $-19379.4(12)$ \\
    $^{36}$Cl & $-29522.01(4)$    & $4299.667(14)$  & $-25222.35(4)$ \\
    $^{36}$S  & $-30664.12(19)$   & --              & $-30664.12(19)$ \\
    \end{tabular}
    \end{ruledtabular}
    \end{table}
%####table was here
%
%
%
%
%
%
% \begin{table*}[t]
%    \caption{\label{tab:IMMEcoefs} Updated coefficients of several variations of the IMME and the quality of fit, %described by chi-squared per degree of freedom. Experimental inputs are given in Table \ref{tab:IMMEinput}.}
%    \begin{ruledtabular}
%    \begin{tabular}{ccccccc}
%     a  &  b  &  c  &  d  &  e  &  $\chi^2/\nu$ & prob. \\
%     \hline
%     $-19378.9 \pm 0.5$ & $-6044.3 \pm 0.8$ & $200.8 \pm 0.3$ & --             & -- & 0.684/2 & 0.710 \\
%     $-19379.6 \pm 1.0$ & $-6043.8 \pm 1.0$ & $201.4 \pm 0.7$ & $-0.3 \pm 0.4$ & -- & 0.082/1 & 0.775 \\
%     $-19379.4 \pm 1.2$ & $-6044.4 \pm 0.9$ & $201.6 \pm 1.8$ & -- & $0.1 \pm 0.4$  & 0.491/1 & 0.483 \\
%     $-19379.4 \pm 1.2$ & $-6043.4 \pm 1.7$ & $200.8 \pm 2.2$ & $-0.4 \pm 0.6$ & $0.2 \pm 0.6$ & -- & --\\
%    \end{tabular}
%    \end{ruledtabular}
%    \end{table*}
    
    \begin{table*}[t]
    \caption{\label{tab:IMMEcoefs} Updated coefficients of several variations of the IMME and the quality of fit, described by chi-squared per degree of freedom and corresponding p-value. Experimental inputs are given in Table \ref{tab:IMMEinput}.}
    \begin{ruledtabular}
    \begin{tabular}{ccccccc}
     a  &  b  &  c  &  d  &  e  &  $\chi^2/\nu$ & p-value \\
     \hline
     $-19378.9 \pm 0.5$ & $-6044.2 \pm 0.8$ & $200.8 \pm 0.3$ & --             & -- & 0.646/2 & 0.724 \\
     $-19379.6 \pm 1.0$ & $-6043.8 \pm 1.0$ & $201.3 \pm 0.7$ & $-0.3 \pm 0.4$ & -- & 0.077/1 & 0.781 \\
     $-19379.4 \pm 1.2$ & $-6044.4 \pm 0.9$ & $201.6 \pm 1.8$ & -- & $-0.1 \pm 0.3$  & 0.463/1 & 0.496 \\
     $-19379.4 \pm 1.2$ & $-6043.5 \pm 1.7$ & $200.8 \pm 2.2$ & $-0.4 \pm 0.6$ & $0.2 \pm 0.6$ & -- & --\\
    \end{tabular}
    \end{ruledtabular}
    \end{table*}

%\section{\label{sec:discussion}Discussion}  
\subsection{\label{subsec:GIMME}The GIMME}  
\begin{figure}[b]
\includegraphics[width=0.475\textwidth]{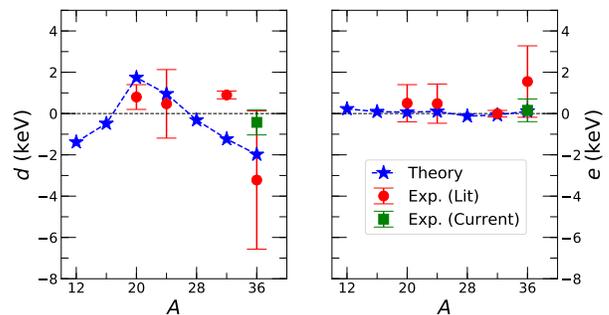}
\caption{\label{fig:DEupdate} $d$ and $e$ coefficients of the $T = 2$ quartic IMME. The star points come from GIMME theory \cite{Dong19}. The circular points are from the experimental literature \cite{MacC14,Glas15} and the square point is the update to the $A=36$ quintet from this work. Those quintets with experimental error bars larger than the figure are omitted. Update to Fig. 3 from Ref. \cite{Dong19}}
\end{figure}

We plot the $T=2$ IMME $d$ and $e$ coefficients in Fig. \ref{fig:DEupdate}, similar to the presentation in Ref. \cite{Dong19}, updated with the present work on the $A=36$ quintet. Our new calculation of the experimental $d$ coefficient is consistent with the IMME ($\delta\sim0.7\sigma$) and less-so with the GIMME prediction ($\delta\sim2.7\sigma$), where $\delta$ is the prediction/experiment difference and $\sigma$ is the one standard deviation experimental error. The experimental $d$ coefficient is between the predictions of the IMME and GIMME, as in the $A=20$ quintet. The $e$ coefficient is consistent with both the IMME and GIMME, but the predictions of the GIMME are small across $A$ and, therefore, similar to the IMME.

Currently, no experimental $e$ coefficient can discriminate between the two predictions, as their difference is beyond an order of magnitude smaller than most of the quintets' experimental error bars. The oscillatory behavior of the $d$ coefficient over $A$ predicted by the GIMME (and strongly evidenced in the $T=3/2$ quartets) appears muted in the $T=2$ cases. The pattern in the GIMME theory papers \cite{Dong18,Dong19} was attributed to nuclear shell structure, with predictions of small $d$, $e$ coefficients in multiplets containing magic or near-magic numbers, so it is peculiar to see only weak experimental evidence in $T=2$.

We don't rule out the GIMME prediction for $A=36$. This rough compatibility highlights the nearby $A=32$ quintet, which departs from both the IMME and GIMME with a positive $d$ coefficient, $d_{exp} = 0.90(19)$ keV \cite{MacC14}. A recent theoretical discussion of the $A=32$ quintet \cite{Sign11} frames this departure from the quadratic IMME as a consequence of isospin mixing. Isospin mixing strength is calculated with the USD, USDA, USDB shell model interactions and each predicts a positive $d$ coefficient ($d_{theory} = 0.28-0.39$ keV), similar to, but smaller than the experimental value. However, mixing strength depends strongly on the precise difference in energy of nearby states capable of mixing. A confounding factor of these calculations is the 100-150 keV rms deviation typical of the empirical interactions; a 100 keV shift in any of the participating states' energy lends itself to a considerable change in the effects predicted due to isospin mixing.

%in their work, there is compromise between reconstructing the excitation energy spectrum in $^{32}$Cl ($T_z = -1$), necessary to correctly calculate state-mixing induced excitation energy shifts, and the measured proton emission strength from the mixed $T=1$ states. 

%\section{\label{sec:outlook}Outlook}
\subsection{\label{subsec:outlook}Outlook}
 Given the agreement between experiment and GIMME prediction at the lowest mass quintets and the tension in the $A = 32, 36$ quintets, it is important to precisely measure the quintets in the middle. In both the $A = 24, 28$ quintets, uncertainty contribution in the $d$ and $e$ coefficients is led by the ground state mass measurements of the $T_z=-2$ members, ($^{24}$Si, $^{28}$S). The mass of $^{24}$Si was recently measured by LEBIT to a precision that may offer discrimination between the IMME and GIMME predictions, and the publication is currently being prepared. A measurement of $^{28}$S will finish the $A = 28$ quintet, giving IMME coefficients from $A = 20 - 36$ in regular steps of $\delta A = 4$. Additionally, paired with recent proton-separation energy measurements \cite{Mukh15}, a high-precision measurement of $^{28}$S would update the mass of $^{29}$Cl and may clarify the source of the large $T=5/2, A=29$ $d$ coefficient ($d = 28(7)$ keV), the largest measured cubic coefficient across all masses and isospin \cite{Brod17}. Meanwhile, further insight into the source of the positive experimental $d$ coefficient in $A = 32$ may explain the apparent breakdown of the higher mass $T = 2$ GIMME predictions. 
 
 In principle, $T = 2$ quintets of mass $A = 40$ and higher may be accessible experimentally. The $A = 40$ quintet currently would benefit from Penning trap mass measurements of both $T_z = -1$ ($^{40}$Sc) and $T_z = -2$ ($^{40}$Ti) members. A large barrier to these measurements is low production cross-sections due to these isotopes' proximity to the proton-drip line. Additionally, the measured half-lives of $T_z = -2$ members are short ($\sim 50$ ms). However, advanced Penning trap techniques like Phase Imaging - Ion Cyclotron Resonance \cite{Elis13,Nest18,Orfo20} will extend the half-life reach of Penning trap mass measurements with rare isotopes and next-generation fragmentation facilities are expected to have favorable beam rates. For most $A = 4n + 2$ quintets ($n$ is an integer), the $T_z = -2$ members are proton-unbound. The exceptions are $A = 22, 26$, however no $T = 2$ IAS have yet been identified in odd-odd $T_z = 0$ nuclei $^{22}$Na and $^{26}$Al.
 
 There is considerable scientific interest in measuring the ground state mass of the above mentioned $T_z = -2$ nuclei to $1$ keV or better for tests of the standard model. The precise measurements of $T = 2$ superallowed $0^+\rightarrow0^+$ decay properties such as those measured in the decay of $^{32}$Ar in Ref. \cite{Bhat08} stand to build upon the comprehensive evaluation of $T = 1$ superallowed $0^+\rightarrow0^+$ decays by Towner and Hardy \cite{Hard15} testing the Conserved Vector Current hypothesis and consequentially providing the most precisely known element in the quark-mixing Cabibbo-Kobayashi-Maskawa matrix. Scalar currents in the weak interaction can be probed by measuring the beta-neutrino angular correlation in the decays of $T = 2$ nuclei and modern programs stand to improve on earlier work done with $^{32}$Ar \cite{Adel99} (TAMUTRAP \cite{Mehl13,Shid19}, WISArD \cite{Asch16,Arau20}). Additionally, one may set limits on electron-sterile neutrino mixing from precision measurements of the beta-particle energy distribution \cite{Schr80,Brym19}. Measurements of $T = 2$ nuclei to the precision required for these tests will provide stronger tests of the GIMME $d$ coefficient, and may be sensitive enough to unambiguously measure finite values for the largest predicted $e$ coefficients.

\section{\label{sec:conclusion}Conclusions}
We have reported the mass excess of neutron-deficient $^{36}$Ca, measured directly for the first time with the Time of Flight - Ion Cyclotron Resonance method in the LEBIT 9.4 T Penning trap at the National Superconducting Cyclotron Laboratory. This completes the third high-precision isobaric quintet with the challenging $T_z = -2$ member ground state mass determined to high-precision with Penning trap mass spectrometry. We interpreted this new result in the context of the Isobaric Multiplet Mass Equation and a new \textit{ab initio} derived calculation that naturally extends the IMME. We find excellent agreement in the $A = 36, T = 2$ isobaric multiplet with the quadratic IMME, and only coarse qualitative agreement with the new GIMME predictions. Tension is emerging between GIMME $T = 2$ predictions and experimental measurements for $A = 32$ and heavier.

\section{\label{sec:acknowledgement}Acknowledgements}

We thank J.M. Dong for sharing the numerical predictions from the GIMME theory to make Fig. \ref{fig:DEupdate}. This work was conducted with the support of Michigan State University, the National Science Foundation under Contracts No. PHY-1565546, PHY-1713857, and PHY-1913554, the U.S. Department of Energy, Office of Science, Office of Nuclear Physics under Award No. DE-SC-0015927 and DE-SC-0016052, and NSERC (Canada) under Contract No. SAPPJ-2018-00028 .

%\nocite{*}

% uncomment for nicely made bib
%\bibliography{36ca}% Produces the bibliography via BibTeX.

\begin{thebibliography}{46}%
\makeatletter
\providecommand \@ifxundefined [1]{%
 \@ifx{#1\undefined}
}%
\providecommand \@ifnum [1]{%
 \ifnum #1\expandafter \@firstoftwo
 \else \expandafter \@secondoftwo
 \fi
}%
\providecommand \@ifx [1]{%
 \ifx #1\expandafter \@firstoftwo
 \else \expandafter \@secondoftwo
 \fi
}%
\providecommand \natexlab [1]{#1}%
\providecommand \enquote  [1]{``#1''}%
\providecommand \bibnamefont  [1]{#1}%
\providecommand \bibfnamefont [1]{#1}%
\providecommand \citenamefont [1]{#1}%
\providecommand \href@noop [0]{\@secondoftwo}%
\providecommand \href [0]{\begingroup \@sanitize@url \@href}%
\providecommand \@href[1]{\@@startlink{#1}\@@href}%
\providecommand \@@href[1]{\endgroup#1\@@endlink}%
\providecommand \@sanitize@url [0]{\catcode `\\12\catcode `\$12\catcode
  `\&12\catcode `\#12\catcode `\^12\catcode `\_12\catcode `\%12\relax}%
\providecommand \@@startlink[1]{}%
\providecommand \@@endlink[0]{}%
\providecommand \url  [0]{\begingroup\@sanitize@url \@url }%
\providecommand \@url [1]{\endgroup\@href {#1}{\urlprefix }}%
\providecommand \urlprefix  [0]{URL }%
\providecommand \Eprint [0]{\href }%
\providecommand \doibase [0]{https://doi.org/}%
\providecommand \selectlanguage [0]{\@gobble}%
\providecommand \bibinfo  [0]{\@secondoftwo}%
\providecommand \bibfield  [0]{\@secondoftwo}%
\providecommand \translation [1]{[#1]}%
\providecommand \BibitemOpen [0]{}%
\providecommand \bibitemStop [0]{}%
\providecommand \bibitemNoStop [0]{.\EOS\space}%
\providecommand \EOS [0]{\spacefactor3000\relax}%
\providecommand \BibitemShut  [1]{\csname bibitem#1\endcsname}%
\let\auto@bib@innerbib\@empty
%</preamble>
\bibitem [{\citenamefont {Heisenberg}(1932)}]{Heis32}%
  \BibitemOpen
  \bibfield  {author} {\bibinfo {author} {\bibfnamefont {W.}~\bibnamefont
  {Heisenberg}},\ }\href@noop {} {\bibfield  {journal} {\bibinfo  {journal}
  {Zeit. Phys.}\ }\textbf {\bibinfo {volume} {77}},\ \bibinfo {pages} {1}
  (\bibinfo {year} {1932})}\BibitemShut {NoStop}%
\bibitem [{\citenamefont {Weinberg}\ and\ \citenamefont
  {Treiman}(1959)}]{Wein59}%
  \BibitemOpen
  \bibfield  {author} {\bibinfo {author} {\bibfnamefont {S.}~\bibnamefont
  {Weinberg}}\ and\ \bibinfo {author} {\bibfnamefont {S.~B.}\ \bibnamefont
  {Treiman}},\ }\href@noop {} {\bibfield  {journal} {\bibinfo  {journal} {Phys.
  Rev.}\ }\textbf {\bibinfo {volume} {116}},\ \bibinfo {pages} {465} (\bibinfo
  {year} {1959})}\BibitemShut {NoStop}%
\bibitem [{\citenamefont {Henley}\ and\ \citenamefont {Lacy}(1969)}]{Henl69}%
  \BibitemOpen
  \bibfield  {author} {\bibinfo {author} {\bibfnamefont {E.~M.}\ \bibnamefont
  {Henley}}\ and\ \bibinfo {author} {\bibfnamefont {C.~E.}\ \bibnamefont
  {Lacy}},\ }\href@noop {} {\bibfield  {journal} {\bibinfo  {journal} {Phys.
  Rev.}\ }\textbf {\bibinfo {volume} {184}},\ \bibinfo {pages} {1228} (\bibinfo
  {year} {1969})}\BibitemShut {NoStop}%
\bibitem [{\citenamefont {J{\"a}necke}(1969)}]{Jane69}%
  \BibitemOpen
  \bibfield  {author} {\bibinfo {author} {\bibfnamefont {J.}~\bibnamefont
  {J{\"a}necke}},\ }\href@noop {} {\bibfield  {journal} {\bibinfo  {journal}
  {Nucl. Phys. A}\ }\textbf {\bibinfo {volume} {128}},\ \bibinfo {pages} {632}
  (\bibinfo {year} {1969})}\BibitemShut {NoStop}%
\bibitem [{\citenamefont {Bertsch}\ and\ \citenamefont
  {Kahana}(1970)}]{Bert70}%
  \BibitemOpen
  \bibfield  {author} {\bibinfo {author} {\bibfnamefont {G.}~\bibnamefont
  {Bertsch}}\ and\ \bibinfo {author} {\bibfnamefont {S.}~\bibnamefont
  {Kahana}},\ }\href@noop {} {\bibfield  {journal} {\bibinfo  {journal} {Phys.
  Let. B}\ }\textbf {\bibinfo {volume} {33}},\ \bibinfo {pages} {193} (\bibinfo
  {year} {1970})}\BibitemShut {NoStop}%
\bibitem [{\citenamefont {Wiringa}\ \emph {et~al.}(1995)\citenamefont
  {Wiringa}, \citenamefont {Stoks},\ and\ \citenamefont {Schiavilla}}]{Wiri95}%
  \BibitemOpen
  \bibfield  {author} {\bibinfo {author} {\bibfnamefont {R.~B.}\ \bibnamefont
  {Wiringa}}, \bibinfo {author} {\bibfnamefont {V.~G.~J.}\ \bibnamefont
  {Stoks}},\ and\ \bibinfo {author} {\bibfnamefont {R.}~\bibnamefont
  {Schiavilla}},\ }\href@noop {} {\bibfield  {journal} {\bibinfo  {journal}
  {Phys. Rev. C}\ }\textbf {\bibinfo {volume} {51}},\ \bibinfo {pages} {38}
  (\bibinfo {year} {1995})}\BibitemShut {NoStop}%
\bibitem [{\citenamefont {Brueckner}\ \emph {et~al.}(1958)\citenamefont
  {Brueckner}, \citenamefont {Gammel},\ and\ \citenamefont
  {Weitzner}}]{Brue58}%
  \BibitemOpen
  \bibfield  {author} {\bibinfo {author} {\bibfnamefont {K.}~\bibnamefont
  {Brueckner}}, \bibinfo {author} {\bibfnamefont {J.}~\bibnamefont {Gammel}},\
  and\ \bibinfo {author} {\bibfnamefont {H.}~\bibnamefont {Weitzner}},\
  }\href@noop {} {\bibfield  {journal} {\bibinfo  {journal} {Phys. Rev.}\
  }\textbf {\bibinfo {volume} {110}},\ \bibinfo {pages} {431} (\bibinfo {year}
  {1958})}\BibitemShut {NoStop}%
\bibitem [{\citenamefont {Dong}\ \emph {et~al.}(2018)\citenamefont {Dong},
  \citenamefont {Zhang}, \citenamefont {Zuo}, \citenamefont {Gu}, \citenamefont
  {Wang},\ and\ \citenamefont {Sun}}]{Dong18}%
  \BibitemOpen
  \bibfield  {author} {\bibinfo {author} {\bibfnamefont {J.~M.}\ \bibnamefont
  {Dong}}, \bibinfo {author} {\bibfnamefont {Y.~H.}\ \bibnamefont {Zhang}},
  \bibinfo {author} {\bibfnamefont {W.}~\bibnamefont {Zuo}}, \bibinfo {author}
  {\bibfnamefont {J.~Z.}\ \bibnamefont {Gu}}, \bibinfo {author} {\bibfnamefont
  {L.~J.}\ \bibnamefont {Wang}},\ and\ \bibinfo {author} {\bibfnamefont
  {Y.}~\bibnamefont {Sun}},\ }\href@noop {} {\bibfield  {journal} {\bibinfo
  {journal} {Phys. Rev. C}\ }\textbf {\bibinfo {volume} {97}},\ \bibinfo
  {pages} {021301(R)} (\bibinfo {year} {2018})}\BibitemShut {NoStop}%
\bibitem [{\citenamefont {Dong}\ \emph {et~al.}(2019)\citenamefont {Dong},
  \citenamefont {Gu}, \citenamefont {Zhang}, \citenamefont {Zuo}, \citenamefont
  {Wang}, \citenamefont {Litvinov},\ and\ \citenamefont {Sun}}]{Dong19}%
  \BibitemOpen
  \bibfield  {author} {\bibinfo {author} {\bibfnamefont {J.~M.}\ \bibnamefont
  {Dong}}, \bibinfo {author} {\bibfnamefont {J.~Z.}\ \bibnamefont {Gu}},
  \bibinfo {author} {\bibfnamefont {Y.~H.}\ \bibnamefont {Zhang}}, \bibinfo
  {author} {\bibfnamefont {W.}~\bibnamefont {Zuo}}, \bibinfo {author}
  {\bibfnamefont {L.~J.}\ \bibnamefont {Wang}}, \bibinfo {author}
  {\bibfnamefont {Y.~A.}\ \bibnamefont {Litvinov}},\ and\ \bibinfo {author}
  {\bibfnamefont {Y.}~\bibnamefont {Sun}},\ }\href@noop {} {\bibfield
  {journal} {\bibinfo  {journal} {Phys. Rev. C}\ }\textbf {\bibinfo {volume}
  {99}},\ \bibinfo {pages} {014319} (\bibinfo {year} {2019})}\BibitemShut
  {NoStop}%
\bibitem [{\citenamefont {MacCormick}\ and\ \citenamefont
  {Audi}(2014)}]{MacC14}%
  \BibitemOpen
  \bibfield  {author} {\bibinfo {author} {\bibfnamefont {M.}~\bibnamefont
  {MacCormick}}\ and\ \bibinfo {author} {\bibfnamefont {G.}~\bibnamefont
  {Audi}},\ }\href@noop {} {\bibfield  {journal} {\bibinfo  {journal} {Nucl.
  Phys. A}\ }\textbf {\bibinfo {volume} {925}},\ \bibinfo {pages} {61}
  (\bibinfo {year} {2014})}\BibitemShut {NoStop}%
\bibitem [{\citenamefont {Wang}\ \emph {et~al.}(2017)\citenamefont {Wang},
  \citenamefont {Audi}, \citenamefont {Kondev}, \citenamefont {Huang},
  \citenamefont {Naimi},\ and\ \citenamefont {Xu}}]{AME2016}%
  \BibitemOpen
  \bibfield  {author} {\bibinfo {author} {\bibfnamefont {M.}~\bibnamefont
  {Wang}}, \bibinfo {author} {\bibfnamefont {G.}~\bibnamefont {Audi}}, \bibinfo
  {author} {\bibfnamefont {F.~G.}\ \bibnamefont {Kondev}}, \bibinfo {author}
  {\bibfnamefont {W.~J.}\ \bibnamefont {Huang}}, \bibinfo {author}
  {\bibfnamefont {S.}~\bibnamefont {Naimi}},\ and\ \bibinfo {author}
  {\bibfnamefont {X.}~\bibnamefont {Xu}},\ }\href@noop {} {\bibfield  {journal}
  {\bibinfo  {journal} {Chin. Phys. C}\ }\textbf {\bibinfo {volume} {41}},\
  \bibinfo {pages} {030003} (\bibinfo {year} {2017})}\BibitemShut {NoStop}%
\bibitem [{\citenamefont {Glassman}\ \emph {et~al.}(2015)\citenamefont
  {Glassman}, \citenamefont {P{\'e}rez-Loureiro}, \citenamefont {Wrede},
  \citenamefont {Allen}, \citenamefont {Bardayan}, \citenamefont {Bennett},
  \citenamefont {Brown}, \citenamefont {Chipps}, \citenamefont {Febbraro},
  \citenamefont {Fry}, \citenamefont {Hall}, \citenamefont {Hall},
  \citenamefont {Liddick}, \citenamefont {O’Malley}, \citenamefont {Ong},
  \citenamefont {Pain}, \citenamefont {Schwartz}, \citenamefont {Shidling},
  \citenamefont {Sims}, \citenamefont {Thompson},\ and\ \citenamefont
  {Zhang}}]{Glas15}%
  \BibitemOpen
  \bibfield  {author} {\bibinfo {author} {\bibfnamefont {B.~E.}\ \bibnamefont
  {Glassman}}, \bibinfo {author} {\bibfnamefont {D.}~\bibnamefont
  {P{\'e}rez-Loureiro}}, \bibinfo {author} {\bibfnamefont {C.}~\bibnamefont
  {Wrede}}, \bibinfo {author} {\bibfnamefont {J.}~\bibnamefont {Allen}},
  \bibinfo {author} {\bibfnamefont {D.~W.}\ \bibnamefont {Bardayan}}, \bibinfo
  {author} {\bibfnamefont {M.~B.}\ \bibnamefont {Bennett}}, \bibinfo {author}
  {\bibfnamefont {B.~A.}\ \bibnamefont {Brown}}, \bibinfo {author}
  {\bibfnamefont {K.~A.}\ \bibnamefont {Chipps}}, \bibinfo {author}
  {\bibfnamefont {M.}~\bibnamefont {Febbraro}}, \bibinfo {author}
  {\bibfnamefont {C.}~\bibnamefont {Fry}}, \bibinfo {author} {\bibfnamefont
  {M.~R.}\ \bibnamefont {Hall}}, \bibinfo {author} {\bibfnamefont
  {O.}~\bibnamefont {Hall}}, \bibinfo {author} {\bibfnamefont {S.~N.}\
  \bibnamefont {Liddick}}, \bibinfo {author} {\bibfnamefont {P.}~\bibnamefont
  {O’Malley}}, \bibinfo {author} {\bibfnamefont {W.}~\bibnamefont {Ong}},
  \bibinfo {author} {\bibfnamefont {S.~D.}\ \bibnamefont {Pain}}, \bibinfo
  {author} {\bibfnamefont {S.~B.}\ \bibnamefont {Schwartz}}, \bibinfo {author}
  {\bibfnamefont {P.}~\bibnamefont {Shidling}}, \bibinfo {author}
  {\bibfnamefont {H.}~\bibnamefont {Sims}}, \bibinfo {author} {\bibfnamefont
  {P.}~\bibnamefont {Thompson}},\ and\ \bibinfo {author} {\bibfnamefont
  {H.}~\bibnamefont {Zhang}},\ }\href@noop {} {\bibfield  {journal} {\bibinfo
  {journal} {Phys. Rev. C}\ }\textbf {\bibinfo {volume} {92}},\ \bibinfo
  {pages} {042501(R)} (\bibinfo {year} {2015})}\BibitemShut {NoStop}%
\bibitem [{\citenamefont {Blaum}(2006)}]{Blau06}%
  \BibitemOpen
  \bibfield  {author} {\bibinfo {author} {\bibfnamefont {K.}~\bibnamefont
  {Blaum}},\ }\href@noop {} {\bibfield  {journal} {\bibinfo  {journal} {Phys.
  Reports}\ }\textbf {\bibinfo {volume} {425}},\ \bibinfo {pages} {1} (\bibinfo
  {year} {2006})}\BibitemShut {NoStop}%
\bibitem [{\citenamefont {Lunney}\ \emph {et~al.}(2003)\citenamefont {Lunney},
  \citenamefont {Pearson},\ and\ \citenamefont {Thibault}}]{Lunn03}%
  \BibitemOpen
  \bibfield  {author} {\bibinfo {author} {\bibfnamefont {D.}~\bibnamefont
  {Lunney}}, \bibinfo {author} {\bibfnamefont {J.~M.}\ \bibnamefont
  {Pearson}},\ and\ \bibinfo {author} {\bibfnamefont {C.}~\bibnamefont
  {Thibault}},\ }\href@noop {} {\bibfield  {journal} {\bibinfo  {journal} {Rev.
  Mod. Phys.}\ }\textbf {\bibinfo {volume} {75}},\ \bibinfo {pages} {1021}
  (\bibinfo {year} {2003})}\BibitemShut {NoStop}%
\bibitem [{\citenamefont {Tribble}\ \emph {et~al.}(1977)\citenamefont
  {Tribble}, \citenamefont {Cossairt},\ and\ \citenamefont
  {Kenefick}}]{Trib77}%
  \BibitemOpen
  \bibfield  {author} {\bibinfo {author} {\bibfnamefont {R.~E.}\ \bibnamefont
  {Tribble}}, \bibinfo {author} {\bibfnamefont {J.~D.}\ \bibnamefont
  {Cossairt}},\ and\ \bibinfo {author} {\bibfnamefont {R.~A.}\ \bibnamefont
  {Kenefick}},\ }\href@noop {} {\bibfield  {journal} {\bibinfo  {journal}
  {Phys. Rev. C}\ }\textbf {\bibinfo {volume} {15}},\ \bibinfo {pages} {2028}
  (\bibinfo {year} {1977})}\BibitemShut {NoStop}%
\bibitem [{\citenamefont {Komives}\ \emph {et~al.}(1998)\citenamefont
  {Komives}, \citenamefont {Garc{\'i}a}, \citenamefont {Peterson}, \citenamefont
  {Bazin}, \citenamefont {Caggiano}, \citenamefont {Sherrill}, \citenamefont
  {Alahari}, \citenamefont {Bacher}, \citenamefont {Lozowski}, \citenamefont
  {Greene},\ and\ \citenamefont {Adelberger}}]{Komi98}%
  \BibitemOpen
  \bibfield  {author} {\bibinfo {author} {\bibfnamefont {A.}~\bibnamefont
  {Komives}}, \bibinfo {author} {\bibfnamefont {A.}~\bibnamefont {Garc{\'i}a}},
  \bibinfo {author} {\bibfnamefont {D.}~\bibnamefont {Peterson}}, \bibinfo
  {author} {\bibfnamefont {D.}~\bibnamefont {Bazin}}, \bibinfo {author}
  {\bibfnamefont {J.}~\bibnamefont {Caggiano}}, \bibinfo {author}
  {\bibfnamefont {B.}~\bibnamefont {Sherrill}}, \bibinfo {author}
  {\bibfnamefont {N.}~\bibnamefont {Alahari}}, \bibinfo {author} {\bibfnamefont
  {A.}~\bibnamefont {Bacher}}, \bibinfo {author} {\bibfnamefont
  {W.}~\bibnamefont {Lozowski}}, \bibinfo {author} {\bibfnamefont
  {J.}~\bibnamefont {Greene}},\ and\ \bibinfo {author} {\bibfnamefont {E.~G.}\
  \bibnamefont {Adelberger}},\ }\bibfield  {title} {\bibinfo {title} {The mass
  of $^{36}$\normalsize{}\textsc{C}\small{}a and the isobaric mass multiplet
  equation},\ }\href@noop {} {\bibfield  {journal} {\bibinfo  {journal}
  {American Physical Society, Division of Nuclear Physics Meeting, October
  28-31, 1998 Santa Fe, New Mexico}\ } (\bibinfo {year} {1998})},\ \bibinfo
  {note} {abstract ID. F3.07}\BibitemShut {NoStop}%
\bibitem [{\citenamefont {Greene}\ \emph {et~al.}(1999)\citenamefont {Greene},
  \citenamefont {Thomas}, \citenamefont {Garc{\'i}a}, \citenamefont {Komives},\
  and\ \citenamefont {Jr.}}]{Gree99}%
  \BibitemOpen
  \bibfield  {author} {\bibinfo {author} {\bibfnamefont {J.~P.}\ \bibnamefont
  {Greene}}, \bibinfo {author} {\bibfnamefont {G.~E.}\ \bibnamefont {Thomas}},
  \bibinfo {author} {\bibfnamefont {A.}~\bibnamefont {Garc{\'i}a}}, \bibinfo
  {author} {\bibfnamefont {A.}~\bibnamefont {Komives}},\ and\ \bibinfo {author}
  {\bibfnamefont {J.~O.~S.}\ \bibnamefont {Jr.}},\ }\href@noop {} {\bibfield
  {journal} {\bibinfo  {journal} {Nucl. Instrum. Methods Phys. Res. A}\
  }\textbf {\bibinfo {volume} {438}},\ \bibinfo {pages} {52} (\bibinfo {year}
  {1999})}\BibitemShut {NoStop}%
\bibitem [{\citenamefont {Caggiano}(1999)}]{Cagg99}%
  \BibitemOpen
  \bibfield  {author} {\bibinfo {author} {\bibfnamefont {J.~A.}\ \bibnamefont
  {Caggiano}},\ }\emph {\bibinfo {title} {Spectroscopy of Exotic Nuclei With
  The S800 Spectrograph}},\ \href@noop {} {\bibinfo {type} {{Ph.D.} thesis}},\
  \bibinfo  {school} {Michigan State University} (\bibinfo {year}
  {1999})\BibitemShut {NoStop}%
\bibitem [{\citenamefont {Morrissey}\ \emph {et~al.}(2003)\citenamefont
  {Morrissey}, \citenamefont {Sherrill}, \citenamefont {Steiner}, \citenamefont
  {Stolz},\ and\ \citenamefont {Wiedenhoever}}]{Morr03}%
  \BibitemOpen
  \bibfield  {author} {\bibinfo {author} {\bibfnamefont {D.~J.}\ \bibnamefont
  {Morrissey}}, \bibinfo {author} {\bibfnamefont {B.~M.}\ \bibnamefont
  {Sherrill}}, \bibinfo {author} {\bibfnamefont {M.}~\bibnamefont {Steiner}},
  \bibinfo {author} {\bibfnamefont {A.}~\bibnamefont {Stolz}},\ and\ \bibinfo
  {author} {\bibfnamefont {I.}~\bibnamefont {Wiedenhoever}},\ }\href@noop {}
  {\bibfield  {journal} {\bibinfo  {journal} {Nucl. Instrum. Methods Phys. Res.
  B}\ }\textbf {\bibinfo {volume} {204}},\ \bibinfo {pages} {90} (\bibinfo
  {year} {2003})}\BibitemShut {NoStop}%
\bibitem [{\citenamefont {Sumithrarachchi}\ \emph {et~al.}(2020)\citenamefont
  {Sumithrarachchi}, \citenamefont {Morrissey}, \citenamefont {Schwarz},
  \citenamefont {Lund}, \citenamefont {Bollen}, \citenamefont {Ringle},
  \citenamefont {Savard},\ and\ \citenamefont {Villari}}]{Sumi20}%
  \BibitemOpen
  \bibfield  {author} {\bibinfo {author} {\bibfnamefont {C.~S.}\ \bibnamefont
  {Sumithrarachchi}}, \bibinfo {author} {\bibfnamefont {D.}~\bibnamefont
  {Morrissey}}, \bibinfo {author} {\bibfnamefont {S.}~\bibnamefont {Schwarz}},
  \bibinfo {author} {\bibfnamefont {K.}~\bibnamefont {Lund}}, \bibinfo {author}
  {\bibfnamefont {G.}~\bibnamefont {Bollen}}, \bibinfo {author} {\bibfnamefont
  {R.}~\bibnamefont {Ringle}}, \bibinfo {author} {\bibfnamefont
  {G.}~\bibnamefont {Savard}},\ and\ \bibinfo {author} {\bibfnamefont
  {A.~C.~C.}\ \bibnamefont {Villari}},\ }\href@noop {} {\bibfield  {journal}
  {\bibinfo  {journal} {Nucl. Instrum. Methods Phys. Res. B}\ }\textbf
  {\bibinfo {volume} {463}},\ \bibinfo {pages} {305–309} (\bibinfo {year}
  {2020})}\BibitemShut {NoStop}%
\bibitem [{\citenamefont {Ringle}\ \emph {et~al.}(2013)\citenamefont {Ringle},
  \citenamefont {Schwarz},\ and\ \citenamefont {Bollen}}]{Ring13}%
  \BibitemOpen
  \bibfield  {author} {\bibinfo {author} {\bibfnamefont {R.}~\bibnamefont
  {Ringle}}, \bibinfo {author} {\bibfnamefont {S.}~\bibnamefont {Schwarz}},\
  and\ \bibinfo {author} {\bibfnamefont {G.}~\bibnamefont {Bollen}},\
  }\href@noop {} {\bibfield  {journal} {\bibinfo  {journal} {Int. J. Mass
  Spectrom.}\ }\textbf {\bibinfo {volume} {349-350}},\ \bibinfo {pages} {87}
  (\bibinfo {year} {2013})}\BibitemShut {NoStop}%
\bibitem [{\citenamefont {Schwarz}\ \emph {et~al.}(2016)\citenamefont
  {Schwarz}, \citenamefont {Bollen}, \citenamefont {Ringle}, \citenamefont
  {Savory},\ and\ \citenamefont {Schury}}]{Schw16}%
  \BibitemOpen
  \bibfield  {author} {\bibinfo {author} {\bibfnamefont {S.}~\bibnamefont
  {Schwarz}}, \bibinfo {author} {\bibfnamefont {G.}~\bibnamefont {Bollen}},
  \bibinfo {author} {\bibfnamefont {R.}~\bibnamefont {Ringle}}, \bibinfo
  {author} {\bibfnamefont {J.}~\bibnamefont {Savory}},\ and\ \bibinfo {author}
  {\bibfnamefont {P.}~\bibnamefont {Schury}},\ }\href@noop {} {\bibfield
  {journal} {\bibinfo  {journal} {Nucl. Instrum. Methods Phys. Res. A}\
  }\textbf {\bibinfo {volume} {816}},\ \bibinfo {pages} {131} (\bibinfo {year}
  {2016})}\BibitemShut {NoStop}%
\bibitem [{\citenamefont {Ringle}\ \emph
  {et~al.}(2007{\natexlab{a}})\citenamefont {Ringle}, \citenamefont {Bollen},
  \citenamefont {Prinke}, \citenamefont {Savory}, \citenamefont {Schury},
  \citenamefont {Schwarz},\ and\ \citenamefont {Sun}}]{Ring07}%
  \BibitemOpen
  \bibfield  {author} {\bibinfo {author} {\bibfnamefont {R.}~\bibnamefont
  {Ringle}}, \bibinfo {author} {\bibfnamefont {G.}~\bibnamefont {Bollen}},
  \bibinfo {author} {\bibfnamefont {A.}~\bibnamefont {Prinke}}, \bibinfo
  {author} {\bibfnamefont {J.}~\bibnamefont {Savory}}, \bibinfo {author}
  {\bibfnamefont {P.}~\bibnamefont {Schury}}, \bibinfo {author} {\bibfnamefont
  {S.}~\bibnamefont {Schwarz}},\ and\ \bibinfo {author} {\bibfnamefont
  {T.}~\bibnamefont {Sun}},\ }\href@noop {} {\bibfield  {journal} {\bibinfo
  {journal} {Int. J. Mass Spectrom.}\ }\textbf {\bibinfo {volume} {263}},\
  \bibinfo {pages} {38} (\bibinfo {year} {2007}{\natexlab{a}})}\BibitemShut
  {NoStop}%
\bibitem [{\citenamefont {Brown}\ and\ \citenamefont
  {Gabrielse}(1982)}]{Brow82}%
  \BibitemOpen
  \bibfield  {author} {\bibinfo {author} {\bibfnamefont {L.~S.}\ \bibnamefont
  {Brown}}\ and\ \bibinfo {author} {\bibfnamefont {G.}~\bibnamefont
  {Gabrielse}},\ }\href@noop {} {\bibfield  {journal} {\bibinfo  {journal}
  {Phys. Rev. A}\ }\textbf {\bibinfo {volume} {25}},\ \bibinfo {pages} {2423}
  (\bibinfo {year} {1982})}\BibitemShut {NoStop}%
\bibitem [{\citenamefont {Redshaw}\ \emph {et~al.}(2013)\citenamefont
  {Redshaw}, \citenamefont {Bollen}, \citenamefont {Bustabad}, \citenamefont
  {Kwiatkowski}, \citenamefont {Lincoln}, \citenamefont {Novario},
  \citenamefont {Ringle}, \citenamefont {Schwarz},\ and\ \citenamefont
  {Valverde}}]{Reds13}%
  \BibitemOpen
  \bibfield  {author} {\bibinfo {author} {\bibfnamefont {M.}~\bibnamefont
  {Redshaw}}, \bibinfo {author} {\bibfnamefont {G.}~\bibnamefont {Bollen}},
  \bibinfo {author} {\bibfnamefont {S.}~\bibnamefont {Bustabad}}, \bibinfo
  {author} {\bibfnamefont {A.~A.}\ \bibnamefont {Kwiatkowski}}, \bibinfo
  {author} {\bibfnamefont {D.~L.}\ \bibnamefont {Lincoln}}, \bibinfo {author}
  {\bibfnamefont {S.~J.}\ \bibnamefont {Novario}}, \bibinfo {author}
  {\bibfnamefont {R.}~\bibnamefont {Ringle}}, \bibinfo {author} {\bibfnamefont
  {S.}~\bibnamefont {Schwarz}},\ and\ \bibinfo {author} {\bibfnamefont {A.~A.}\
  \bibnamefont {Valverde}},\ }\href@noop {} {\bibfield  {journal} {\bibinfo
  {journal} {Nucl. Instrum. Methods Phys. Res. B}\ }\textbf {\bibinfo {volume}
  {317}},\ \bibinfo {pages} {510–516} (\bibinfo {year} {2013})}\BibitemShut
  {NoStop}%
\bibitem [{\citenamefont {Kwiatkowski}\ \emph {et~al.}(2015)\citenamefont
  {Kwiatkowski}, \citenamefont {Bollen}, \citenamefont {M.Redshaw},
  \citenamefont {Ringle},\ and\ \citenamefont {Schwarz}}]{Kwia15}%
  \BibitemOpen
  \bibfield  {author} {\bibinfo {author} {\bibfnamefont {A.~A.}\ \bibnamefont
  {Kwiatkowski}}, \bibinfo {author} {\bibfnamefont {G.}~\bibnamefont {Bollen}},
  \bibinfo {author} {\bibnamefont {M.Redshaw}}, \bibinfo {author}
  {\bibfnamefont {R.}~\bibnamefont {Ringle}},\ and\ \bibinfo {author}
  {\bibfnamefont {S.}~\bibnamefont {Schwarz}},\ }\href@noop {} {\bibfield
  {journal} {\bibinfo  {journal} {Int. J. Mass Spectrom.}\ }\textbf {\bibinfo
  {volume} {379}},\ \bibinfo {pages} {9} (\bibinfo {year} {2015})}\BibitemShut
  {NoStop}%
\bibitem [{\citenamefont {Schury}\ \emph {et~al.}(2006)\citenamefont {Schury},
  \citenamefont {Bollen}, \citenamefont {Block}, \citenamefont {Morrissey},
  \citenamefont {Ringle}, \citenamefont {Prinke}, \citenamefont {Savory},
  \citenamefont {Schwarz},\ and\ \citenamefont {Sun}}]{Schu06}%
  \BibitemOpen
  \bibfield  {author} {\bibinfo {author} {\bibfnamefont {P.}~\bibnamefont
  {Schury}}, \bibinfo {author} {\bibfnamefont {G.}~\bibnamefont {Bollen}},
  \bibinfo {author} {\bibfnamefont {M.}~\bibnamefont {Block}}, \bibinfo
  {author} {\bibfnamefont {D.~J.}\ \bibnamefont {Morrissey}}, \bibinfo {author}
  {\bibfnamefont {R.}~\bibnamefont {Ringle}}, \bibinfo {author} {\bibfnamefont
  {A.}~\bibnamefont {Prinke}}, \bibinfo {author} {\bibfnamefont
  {J.}~\bibnamefont {Savory}}, \bibinfo {author} {\bibfnamefont
  {S.}~\bibnamefont {Schwarz}},\ and\ \bibinfo {author} {\bibfnamefont
  {T.}~\bibnamefont {Sun}},\ }\href@noop {} {\bibfield  {journal} {\bibinfo
  {journal} {Hyp. Int.}\ }\textbf {\bibinfo {volume} {173}},\ \bibinfo {pages}
  {165} (\bibinfo {year} {2006})}\BibitemShut {NoStop}%
\bibitem [{\citenamefont {K{\"o}nig}\ \emph {et~al.}(1995)\citenamefont
  {K{\"o}nig}, \citenamefont {Bollen}, \citenamefont {Kluge}, \citenamefont
  {Otto},\ and\ \citenamefont {Szerypo}}]{Koni95}%
  \BibitemOpen
  \bibfield  {author} {\bibinfo {author} {\bibfnamefont {M.}~\bibnamefont
  {K{\"o}nig}}, \bibinfo {author} {\bibfnamefont {G.}~\bibnamefont {Bollen}},
  \bibinfo {author} {\bibfnamefont {H.-J.}\ \bibnamefont {Kluge}}, \bibinfo
  {author} {\bibfnamefont {T.}~\bibnamefont {Otto}},\ and\ \bibinfo {author}
  {\bibfnamefont {J.}~\bibnamefont {Szerypo}},\ }\href@noop {} {\bibfield
  {journal} {\bibinfo  {journal} {Int. J. Mass Spectrom. Ion Proc.}\ }\textbf
  {\bibinfo {volume} {142}},\ \bibinfo {pages} {95} (\bibinfo {year}
  {1995})}\BibitemShut {NoStop}%
\bibitem [{\citenamefont {Gulyuz}\ \emph {et~al.}(2015)\citenamefont {Gulyuz},
  \citenamefont {Ariche}, \citenamefont {Bollen}, \citenamefont {Bustabad},
  \citenamefont {Eibach}, \citenamefont {Izzo}, \citenamefont {Novario},
  \citenamefont {Redshaw}, \citenamefont {Ringle}, \citenamefont {Sandler},
  \citenamefont {Schwarz},\ and\ \citenamefont {Valverde}}]{Guly15}%
  \BibitemOpen
  \bibfield  {author} {\bibinfo {author} {\bibfnamefont {K.}~\bibnamefont
  {Gulyuz}}, \bibinfo {author} {\bibfnamefont {J.}~\bibnamefont {Ariche}},
  \bibinfo {author} {\bibfnamefont {G.}~\bibnamefont {Bollen}}, \bibinfo
  {author} {\bibfnamefont {S.}~\bibnamefont {Bustabad}}, \bibinfo {author}
  {\bibfnamefont {M.}~\bibnamefont {Eibach}}, \bibinfo {author} {\bibfnamefont
  {C.}~\bibnamefont {Izzo}}, \bibinfo {author} {\bibfnamefont {S.~J.}\
  \bibnamefont {Novario}}, \bibinfo {author} {\bibfnamefont {M.}~\bibnamefont
  {Redshaw}}, \bibinfo {author} {\bibfnamefont {R.}~\bibnamefont {Ringle}},
  \bibinfo {author} {\bibfnamefont {R.}~\bibnamefont {Sandler}}, \bibinfo
  {author} {\bibfnamefont {S.}~\bibnamefont {Schwarz}},\ and\ \bibinfo {author}
  {\bibfnamefont {A.~A.}\ \bibnamefont {Valverde}},\ }\href@noop {} {\bibfield
  {journal} {\bibinfo  {journal} {Phys. Rev. C}\ }\textbf {\bibinfo {volume}
  {91}},\ \bibinfo {pages} {055501} (\bibinfo {year} {2015})}\BibitemShut
  {NoStop}%
\bibitem [{\citenamefont {Ringle}\ \emph
  {et~al.}(2007{\natexlab{b}})\citenamefont {Ringle}, \citenamefont {Sun},
  \citenamefont {Bollen}, \citenamefont {Davies}, \citenamefont {Facina},
  \citenamefont {Huikari}, \citenamefont {Kwan}, \citenamefont {Morrissey},
  \citenamefont {Prinke}, \citenamefont {Savory}, \citenamefont {Schury},
  \citenamefont {Schwarz},\ and\ \citenamefont {Sumithrarachchi}}]{Ring07PRC}%
  \BibitemOpen
  \bibfield  {author} {\bibinfo {author} {\bibfnamefont {R.}~\bibnamefont
  {Ringle}}, \bibinfo {author} {\bibfnamefont {T.}~\bibnamefont {Sun}},
  \bibinfo {author} {\bibfnamefont {G.}~\bibnamefont {Bollen}}, \bibinfo
  {author} {\bibfnamefont {D.}~\bibnamefont {Davies}}, \bibinfo {author}
  {\bibfnamefont {M.}~\bibnamefont {Facina}}, \bibinfo {author} {\bibfnamefont
  {J.}~\bibnamefont {Huikari}}, \bibinfo {author} {\bibfnamefont
  {E.}~\bibnamefont {Kwan}}, \bibinfo {author} {\bibfnamefont {D.~J.}\
  \bibnamefont {Morrissey}}, \bibinfo {author} {\bibfnamefont {A.}~\bibnamefont
  {Prinke}}, \bibinfo {author} {\bibfnamefont {J.}~\bibnamefont {Savory}},
  \bibinfo {author} {\bibfnamefont {P.}~\bibnamefont {Schury}}, \bibinfo
  {author} {\bibfnamefont {S.}~\bibnamefont {Schwarz}},\ and\ \bibinfo {author}
  {\bibfnamefont {C.~S.}\ \bibnamefont {Sumithrarachchi}},\ }\href@noop {}
  {\bibfield  {journal} {\bibinfo  {journal} {Phys. Rev. C}\ }\textbf {\bibinfo
  {volume} {75}},\ \bibinfo {pages} {055503} (\bibinfo {year}
  {2007}{\natexlab{b}})}\BibitemShut {NoStop}%
\bibitem [{\citenamefont {Birge}(1932)}]{Birg32}%
  \BibitemOpen
  \bibfield  {author} {\bibinfo {author} {\bibfnamefont {R.~T.}\ \bibnamefont
  {Birge}},\ }\href@noop {} {\bibfield  {journal} {\bibinfo  {journal} {Phys.
  Rev.}\ }\textbf {\bibinfo {volume} {40}},\ \bibinfo {pages} {207} (\bibinfo
  {year} {1932})}\BibitemShut {NoStop}%
\bibitem [{\citenamefont {Signoracci}\ and\ \citenamefont
  {Brown}(2011)}]{Sign11}%
  \BibitemOpen
  \bibfield  {author} {\bibinfo {author} {\bibfnamefont {A.}~\bibnamefont
  {Signoracci}}\ and\ \bibinfo {author} {\bibfnamefont {B.~A.}\ \bibnamefont
  {Brown}},\ }\href@noop {} {\bibfield  {journal} {\bibinfo  {journal} {Phys.
  Rev. C}\ }\textbf {\bibinfo {volume} {84}},\ \bibinfo {pages} {031301(R)}
  (\bibinfo {year} {2011})}\BibitemShut {NoStop}%
\bibitem [{\citenamefont {Mukha}\ \emph {et~al.}(2015)\citenamefont {Mukha},
  \citenamefont {Grigorenko}, \citenamefont {Xu}, \citenamefont {Acosta},
  \citenamefont {Casarejos}, \citenamefont {Ciemny}, \citenamefont {Dominik},
  \citenamefont {Du{\'e}nas-D{\'i}az}, \citenamefont {Dunin}, \citenamefont
  {Espino}, \citenamefont {Estrad{\'e}}, \citenamefont {Farinon}, \citenamefont
  {Fomichev}, \citenamefont {Geissel}, \citenamefont {Golubkova}, \citenamefont
  {Gorshkov}, \citenamefont {Janas}, \citenamefont {Kami{\'n}ski},
  \citenamefont {Kiselev}, \citenamefont {Kn{\"o}bel}, \citenamefont {Krupko},
  \citenamefont {Kuich}, \citenamefont {Litvinov}, \citenamefont
  {Marquinez-Dur{\'a}n}, \citenamefont {Martel}, \citenamefont {Mazzocchi},
  \citenamefont {Nociforo}, \citenamefont {Ord{\'u}z}, \citenamefont
  {Pf{\"u}tzner}, \citenamefont {Pietri}, \citenamefont {Pomorski},
  \citenamefont {Prochazka}, \citenamefont {Rymzhanova}, \citenamefont
  {S{\'a}nchez-Ben{\'i}tez}, \citenamefont {Scheidenberger}, \citenamefont
  {Sharov}, \citenamefont {Simon}, \citenamefont {Sitar}, \citenamefont
  {Slepnev}, \citenamefont {Stanoiu}, \citenamefont {Strmen}, \citenamefont
  {Szarka}, \citenamefont {Takechi}, \citenamefont {Tanaka}, \citenamefont
  {Weick}, \citenamefont {Winkler}, \citenamefont {Winfield},\ and\
  \citenamefont {Zhukov}}]{Mukh15}%
  \BibitemOpen
  \bibfield  {author} {\bibinfo {author} {\bibfnamefont {I.}~\bibnamefont
  {Mukha}}, \bibinfo {author} {\bibfnamefont {L.~V.}\ \bibnamefont
  {Grigorenko}}, \bibinfo {author} {\bibfnamefont {X.}~\bibnamefont {Xu}},
  \bibinfo {author} {\bibfnamefont {L.}~\bibnamefont {Acosta}}, \bibinfo
  {author} {\bibfnamefont {E.}~\bibnamefont {Casarejos}}, \bibinfo {author}
  {\bibfnamefont {A.~A.}\ \bibnamefont {Ciemny}}, \bibinfo {author}
  {\bibfnamefont {W.}~\bibnamefont {Dominik}}, \bibinfo {author} {\bibfnamefont
  {J.}~\bibnamefont {Du{\'e}nas-D{\'i}az}}, \bibinfo {author} {\bibfnamefont
  {V.}~\bibnamefont {Dunin}}, \bibinfo {author} {\bibfnamefont {J.~M.}\
  \bibnamefont {Espino}}, \bibinfo {author} {\bibfnamefont {A.}~\bibnamefont
  {Estrad{\'e}}}, \bibinfo {author} {\bibfnamefont {F.}~\bibnamefont
  {Farinon}}, \bibinfo {author} {\bibfnamefont {A.}~\bibnamefont {Fomichev}},
  \bibinfo {author} {\bibfnamefont {H.}~\bibnamefont {Geissel}}, \bibinfo
  {author} {\bibfnamefont {T.~A.}\ \bibnamefont {Golubkova}}, \bibinfo {author}
  {\bibfnamefont {A.}~\bibnamefont {Gorshkov}}, \bibinfo {author}
  {\bibfnamefont {Z.}~\bibnamefont {Janas}}, \bibinfo {author} {\bibfnamefont
  {G.}~\bibnamefont {Kami{\'n}ski}}, \bibinfo {author} {\bibfnamefont
  {O.}~\bibnamefont {Kiselev}}, \bibinfo {author} {\bibfnamefont
  {R.}~\bibnamefont {Kn{\"o}bel}}, \bibinfo {author} {\bibfnamefont
  {S.}~\bibnamefont {Krupko}}, \bibinfo {author} {\bibfnamefont
  {M.}~\bibnamefont {Kuich}}, \bibinfo {author} {\bibfnamefont {Y.~A.}\
  \bibnamefont {Litvinov}}, \bibinfo {author} {\bibfnamefont {G.}~\bibnamefont
  {Marquinez-Dur{\'a}n}}, \bibinfo {author} {\bibfnamefont {I.}~\bibnamefont
  {Martel}}, \bibinfo {author} {\bibfnamefont {C.}~\bibnamefont {Mazzocchi}},
  \bibinfo {author} {\bibfnamefont {C.}~\bibnamefont {Nociforo}}, \bibinfo
  {author} {\bibfnamefont {A.~K.}\ \bibnamefont {Ord{\'u}z}}, \bibinfo {author}
  {\bibfnamefont {M.}~\bibnamefont {Pf{\"u}tzner}}, \bibinfo {author}
  {\bibfnamefont {S.}~\bibnamefont {Pietri}}, \bibinfo {author} {\bibfnamefont
  {M.}~\bibnamefont {Pomorski}}, \bibinfo {author} {\bibfnamefont
  {A.}~\bibnamefont {Prochazka}}, \bibinfo {author} {\bibfnamefont
  {S.}~\bibnamefont {Rymzhanova}}, \bibinfo {author} {\bibfnamefont {A.~M.}\
  \bibnamefont {S{\'a}nchez-Ben{\'i}tez}}, \bibinfo {author} {\bibfnamefont
  {C.}~\bibnamefont {Scheidenberger}}, \bibinfo {author} {\bibfnamefont
  {P.}~\bibnamefont {Sharov}}, \bibinfo {author} {\bibfnamefont
  {H.}~\bibnamefont {Simon}}, \bibinfo {author} {\bibfnamefont
  {B.}~\bibnamefont {Sitar}}, \bibinfo {author} {\bibfnamefont
  {R.}~\bibnamefont {Slepnev}}, \bibinfo {author} {\bibfnamefont
  {M.}~\bibnamefont {Stanoiu}}, \bibinfo {author} {\bibfnamefont
  {P.}~\bibnamefont {Strmen}}, \bibinfo {author} {\bibfnamefont
  {I.}~\bibnamefont {Szarka}}, \bibinfo {author} {\bibfnamefont
  {M.}~\bibnamefont {Takechi}}, \bibinfo {author} {\bibfnamefont {Y.~K.}\
  \bibnamefont {Tanaka}}, \bibinfo {author} {\bibfnamefont {H.}~\bibnamefont
  {Weick}}, \bibinfo {author} {\bibfnamefont {M.}~\bibnamefont {Winkler}},
  \bibinfo {author} {\bibfnamefont {J.~S.}\ \bibnamefont {Winfield}},\ and\
  \bibinfo {author} {\bibfnamefont {M.~V.}\ \bibnamefont {Zhukov}},\
  }\href@noop {} {\bibfield  {journal} {\bibinfo  {journal} {Phys. Rev. Let.}\
  }\textbf {\bibinfo {volume} {115}},\ \bibinfo {pages} {202501} (\bibinfo
  {year} {2015})}\BibitemShut {NoStop}%
\bibitem [{\citenamefont {Brodeur}\ \emph {et~al.}(2017)\citenamefont
  {Brodeur}, \citenamefont {Kwiatkowski}, \citenamefont {Drozdowski},
  \citenamefont {Andreoiu}, \citenamefont {Burdette}, \citenamefont
  {Chaudhuri}, \citenamefont {Chowdhury}, \citenamefont {Gallant},
  \citenamefont {Grossheim}, \citenamefont {Gwinner}, \citenamefont {Heggen},
  \citenamefont {Holt}, \citenamefont {Klawitter}, \citenamefont {Lassen},
  \citenamefont {Leach}, \citenamefont {Lennarz}, \citenamefont {Nicoloff},
  \citenamefont {Raeder}, \citenamefont {Schultz}, \citenamefont {Stroberg},
  \citenamefont {Teigelhöfer}, \citenamefont {Thompson}, \citenamefont
  {Wieser},\ and\ \citenamefont {Dilling}}]{Brod17}%
  \BibitemOpen
  \bibfield  {author} {\bibinfo {author} {\bibfnamefont {M.}~\bibnamefont
  {Brodeur}}, \bibinfo {author} {\bibfnamefont {A.~A.}\ \bibnamefont
  {Kwiatkowski}}, \bibinfo {author} {\bibfnamefont {O.~M.}\ \bibnamefont
  {Drozdowski}}, \bibinfo {author} {\bibfnamefont {C.}~\bibnamefont
  {Andreoiu}}, \bibinfo {author} {\bibfnamefont {D.}~\bibnamefont {Burdette}},
  \bibinfo {author} {\bibfnamefont {A.}~\bibnamefont {Chaudhuri}}, \bibinfo
  {author} {\bibfnamefont {U.}~\bibnamefont {Chowdhury}}, \bibinfo {author}
  {\bibfnamefont {A.~T.}\ \bibnamefont {Gallant}}, \bibinfo {author}
  {\bibfnamefont {A.}~\bibnamefont {Grossheim}}, \bibinfo {author}
  {\bibfnamefont {G.}~\bibnamefont {Gwinner}}, \bibinfo {author} {\bibfnamefont
  {H.}~\bibnamefont {Heggen}}, \bibinfo {author} {\bibfnamefont {J.~D.}\
  \bibnamefont {Holt}}, \bibinfo {author} {\bibfnamefont {R.}~\bibnamefont
  {Klawitter}}, \bibinfo {author} {\bibfnamefont {J.}~\bibnamefont {Lassen}},
  \bibinfo {author} {\bibfnamefont {K.~G.}\ \bibnamefont {Leach}}, \bibinfo
  {author} {\bibfnamefont {A.}~\bibnamefont {Lennarz}}, \bibinfo {author}
  {\bibfnamefont {C.}~\bibnamefont {Nicoloff}}, \bibinfo {author}
  {\bibfnamefont {S.}~\bibnamefont {Raeder}}, \bibinfo {author} {\bibfnamefont
  {B.~E.}\ \bibnamefont {Schultz}}, \bibinfo {author} {\bibfnamefont {S.~R.}\
  \bibnamefont {Stroberg}}, \bibinfo {author} {\bibfnamefont {A.}~\bibnamefont
  {Teigelhöfer}}, \bibinfo {author} {\bibfnamefont {R.}~\bibnamefont
  {Thompson}}, \bibinfo {author} {\bibfnamefont {M.}~\bibnamefont {Wieser}},\
  and\ \bibinfo {author} {\bibfnamefont {J.}~\bibnamefont {Dilling}},\
  }\href@noop {} {\bibfield  {journal} {\bibinfo  {journal} {Phys. Rev. C}\
  }\textbf {\bibinfo {volume} {96}},\ \bibinfo {pages} {034316} (\bibinfo
  {year} {2017})}\BibitemShut {NoStop}%
\bibitem [{\citenamefont {Eliseev}\ \emph {et~al.}(2013)\citenamefont
  {Eliseev}, \citenamefont {Blaum}, \citenamefont {Block}, \citenamefont
  {Droese}, \citenamefont {Goncharov}, \citenamefont {Ramirez}, \citenamefont
  {Nesterenko}, \citenamefont {Novikov},\ and\ \citenamefont
  {Schweikhard}}]{Elis13}%
  \BibitemOpen
  \bibfield  {author} {\bibinfo {author} {\bibfnamefont {S.}~\bibnamefont
  {Eliseev}}, \bibinfo {author} {\bibfnamefont {K.}~\bibnamefont {Blaum}},
  \bibinfo {author} {\bibfnamefont {M.}~\bibnamefont {Block}}, \bibinfo
  {author} {\bibfnamefont {C.}~\bibnamefont {Droese}}, \bibinfo {author}
  {\bibfnamefont {M.}~\bibnamefont {Goncharov}}, \bibinfo {author}
  {\bibfnamefont {E.~M.}\ \bibnamefont {Ramirez}}, \bibinfo {author}
  {\bibfnamefont {D.~A.}\ \bibnamefont {Nesterenko}}, \bibinfo {author}
  {\bibfnamefont {Y.~N.}\ \bibnamefont {Novikov}},\ and\ \bibinfo {author}
  {\bibfnamefont {L.}~\bibnamefont {Schweikhard}},\ }\href@noop {} {\bibfield
  {journal} {\bibinfo  {journal} {Phys. Rev. Let.}\ }\textbf {\bibinfo {volume}
  {110}},\ \bibinfo {pages} {082501} (\bibinfo {year} {2013})}\BibitemShut
  {NoStop}%
\bibitem [{\citenamefont {Nesterenkoa}\ \emph {et~al.}(2018)\citenamefont
  {Nesterenkoa}, \citenamefont {Eronen}, \citenamefont {Kankainen},
  \citenamefont {Canete}, \citenamefont {Jokinen}, \citenamefont {Moore},
  \citenamefont {Penttil{\"a}}, \citenamefont {Rinta-Antila}, \citenamefont
  {de~Roubin},\ and\ \citenamefont {Vilen}}]{Nest18}%
  \BibitemOpen
  \bibfield  {author} {\bibinfo {author} {\bibfnamefont {D.~A.}\ \bibnamefont
  {Nesterenkoa}}, \bibinfo {author} {\bibfnamefont {T.}~\bibnamefont {Eronen}},
  \bibinfo {author} {\bibfnamefont {A.}~\bibnamefont {Kankainen}}, \bibinfo
  {author} {\bibfnamefont {L.}~\bibnamefont {Canete}}, \bibinfo {author}
  {\bibfnamefont {A.}~\bibnamefont {Jokinen}}, \bibinfo {author} {\bibfnamefont
  {I.~D.}\ \bibnamefont {Moore}}, \bibinfo {author} {\bibfnamefont
  {H.}~\bibnamefont {Penttil{\"a}}}, \bibinfo {author} {\bibfnamefont
  {S.}~\bibnamefont {Rinta-Antila}}, \bibinfo {author} {\bibfnamefont
  {A.}~\bibnamefont {de~Roubin}},\ and\ \bibinfo {author} {\bibfnamefont
  {M.}~\bibnamefont {Vilen}},\ }\href@noop {} {\bibfield  {journal} {\bibinfo
  {journal} {Eur. Phys. J. A}\ }\textbf {\bibinfo {volume} {54}},\ \bibinfo
  {pages} {154} (\bibinfo {year} {2018})}\BibitemShut {NoStop}%
\bibitem [{\citenamefont {Orford}\ \emph {et~al.}(2020)\citenamefont {Orford},
  \citenamefont {Clark}, \citenamefont {Savard}, \citenamefont {Aprahamian},
  \citenamefont {Buchinger}, \citenamefont {Burkey}, \citenamefont {Gorelov},
  \citenamefont {Klimes}, \citenamefont {Morgan}, \citenamefont {Nystrom},
  \citenamefont {Porter}, \citenamefont {Ray},\ and\ \citenamefont
  {Sharma}}]{Orfo20}%
  \BibitemOpen
  \bibfield  {author} {\bibinfo {author} {\bibfnamefont {R.}~\bibnamefont
  {Orford}}, \bibinfo {author} {\bibfnamefont {J.~A.}\ \bibnamefont {Clark}},
  \bibinfo {author} {\bibfnamefont {G.}~\bibnamefont {Savard}}, \bibinfo
  {author} {\bibfnamefont {A.}~\bibnamefont {Aprahamian}}, \bibinfo {author}
  {\bibfnamefont {F.}~\bibnamefont {Buchinger}}, \bibinfo {author}
  {\bibfnamefont {M.~T.}\ \bibnamefont {Burkey}}, \bibinfo {author}
  {\bibfnamefont {D.~A.}\ \bibnamefont {Gorelov}}, \bibinfo {author}
  {\bibfnamefont {J.~W.}\ \bibnamefont {Klimes}}, \bibinfo {author}
  {\bibfnamefont {G.~E.}\ \bibnamefont {Morgan}}, \bibinfo {author}
  {\bibfnamefont {A.}~\bibnamefont {Nystrom}}, \bibinfo {author} {\bibfnamefont
  {W.~S.}\ \bibnamefont {Porter}}, \bibinfo {author} {\bibfnamefont
  {D.}~\bibnamefont {Ray}},\ and\ \bibinfo {author} {\bibfnamefont {K.~S.}\
  \bibnamefont {Sharma}},\ }\href@noop {} {\bibfield  {journal} {\bibinfo
  {journal} {Nucl. Instrum. Methods Phys. Res. B}\ }\textbf {\bibinfo {volume}
  {463}},\ \bibinfo {pages} {491} (\bibinfo {year} {2020})}\BibitemShut
  {NoStop}%
\bibitem [{\citenamefont {Bhattacharya}\ \emph {et~al.}(2008)\citenamefont
  {Bhattacharya}, \citenamefont {Melconian}, \citenamefont {Komives},
  \citenamefont {Triambak}, \citenamefont {Garc{\'i}a}, \citenamefont
  {Adelberger}, \citenamefont {Brown}, \citenamefont {Cooper}, \citenamefont
  {Glasmacher}, \citenamefont {Guimaraes}, \citenamefont {Mantica},
  \citenamefont {Oros-Peusquens}, \citenamefont {Prisciandaro}, \citenamefont
  {Steiner}, \citenamefont {Swanson}, \citenamefont {Tabor}, ,\ and\
  \citenamefont {Wiedeking}}]{Bhat08}%
  \BibitemOpen
  \bibfield  {author} {\bibinfo {author} {\bibfnamefont {M.}~\bibnamefont
  {Bhattacharya}}, \bibinfo {author} {\bibfnamefont {D.}~\bibnamefont
  {Melconian}}, \bibinfo {author} {\bibfnamefont {A.}~\bibnamefont {Komives}},
  \bibinfo {author} {\bibfnamefont {S.}~\bibnamefont {Triambak}}, \bibinfo
  {author} {\bibfnamefont {A.}~\bibnamefont {Garc{\'i}a}}, \bibinfo {author}
  {\bibfnamefont {E.~G.}\ \bibnamefont {Adelberger}}, \bibinfo {author}
  {\bibfnamefont {B.~A.}\ \bibnamefont {Brown}}, \bibinfo {author}
  {\bibfnamefont {M.~W.}\ \bibnamefont {Cooper}}, \bibinfo {author}
  {\bibfnamefont {T.}~\bibnamefont {Glasmacher}}, \bibinfo {author}
  {\bibfnamefont {V.}~\bibnamefont {Guimaraes}}, \bibinfo {author}
  {\bibfnamefont {P.~F.}\ \bibnamefont {Mantica}}, \bibinfo {author}
  {\bibfnamefont {A.~M.}\ \bibnamefont {Oros-Peusquens}}, \bibinfo {author}
  {\bibfnamefont {J.~I.}\ \bibnamefont {Prisciandaro}}, \bibinfo {author}
  {\bibfnamefont {M.}~\bibnamefont {Steiner}}, \bibinfo {author} {\bibfnamefont
  {H.~E.}\ \bibnamefont {Swanson}}, \bibinfo {author} {\bibfnamefont {S.~L.}\
  \bibnamefont {Tabor}}, ,\ and\ \bibinfo {author} {\bibfnamefont
  {M.}~\bibnamefont {Wiedeking}},\ }\href@noop {} {\bibfield  {journal}
  {\bibinfo  {journal} {Phys. Rev. C}\ }\textbf {\bibinfo {volume} {77}},\
  \bibinfo {pages} {065503} (\bibinfo {year} {2008})}\BibitemShut {NoStop}%
\bibitem [{\citenamefont {Hardy}\ and\ \citenamefont {Towner}(2015)}]{Hard15}%
  \BibitemOpen
  \bibfield  {author} {\bibinfo {author} {\bibfnamefont {J.~C.}\ \bibnamefont
  {Hardy}}\ and\ \bibinfo {author} {\bibfnamefont {I.~S.}\ \bibnamefont
  {Towner}},\ }\href@noop {} {\bibfield  {journal} {\bibinfo  {journal} {Phys.
  Rev. C}\ }\textbf {\bibinfo {volume} {91}},\ \bibinfo {pages} {025501}
  (\bibinfo {year} {2015})}\BibitemShut {NoStop}%
\bibitem [{\citenamefont {Adelberger}\ \emph {et~al.}(1999)\citenamefont
  {Adelberger}, \citenamefont {Ortiz}, \citenamefont {Garc{\'i}a}, \citenamefont
  {Swanson}, \citenamefont {Beck}, \citenamefont {Tengblad}, \citenamefont
  {Borge}, \citenamefont {Martel}, \citenamefont {Bichsel},\ and\ \citenamefont
  {the ISOLDE~Collaboration}}]{Adel99}%
  \BibitemOpen
  \bibfield  {author} {\bibinfo {author} {\bibfnamefont {E.~G.}\ \bibnamefont
  {Adelberger}}, \bibinfo {author} {\bibfnamefont {C.}~\bibnamefont {Ortiz}},
  \bibinfo {author} {\bibfnamefont {A.}~\bibnamefont {Garc{\'i}a}}, \bibinfo
  {author} {\bibfnamefont {H.}~\bibnamefont {Swanson}}, \bibinfo {author}
  {\bibfnamefont {M.}~\bibnamefont {Beck}}, \bibinfo {author} {\bibfnamefont
  {O.}~\bibnamefont {Tengblad}}, \bibinfo {author} {\bibfnamefont {M.~J.~G.}\
  \bibnamefont {Borge}}, \bibinfo {author} {\bibfnamefont {I.}~\bibnamefont
  {Martel}}, \bibinfo {author} {\bibfnamefont {H.}~\bibnamefont {Bichsel}},\
  and\ \bibinfo {author} {\bibnamefont {the ISOLDE~Collaboration}},\
  }\href@noop {} {\bibfield  {journal} {\bibinfo  {journal} {Phys. Rev. Let.}\
  }\textbf {\bibinfo {volume} {83}},\ \bibinfo {pages} {1299} (\bibinfo {year}
  {1999})}\BibitemShut {NoStop}%
\bibitem [{\citenamefont {Mehlman}\ \emph {et~al.}(2013)\citenamefont
  {Mehlman}, \citenamefont {Shidling}, \citenamefont {Behling}, \citenamefont
  {Clark}, \citenamefont {Fenker},\ and\ \citenamefont {Melconian}}]{Mehl13}%
  \BibitemOpen
  \bibfield  {author} {\bibinfo {author} {\bibfnamefont {M.}~\bibnamefont
  {Mehlman}}, \bibinfo {author} {\bibfnamefont {P.~D.}\ \bibnamefont
  {Shidling}}, \bibinfo {author} {\bibfnamefont {S.}~\bibnamefont {Behling}},
  \bibinfo {author} {\bibfnamefont {L.~G.}\ \bibnamefont {Clark}}, \bibinfo
  {author} {\bibfnamefont {B.}~\bibnamefont {Fenker}},\ and\ \bibinfo {author}
  {\bibfnamefont {D.}~\bibnamefont {Melconian}},\ }\href@noop {} {\bibfield
  {journal} {\bibinfo  {journal} {Nucl. Instrum. Methods Phys. Res. A}\
  }\textbf {\bibinfo {volume} {712}},\ \bibinfo {pages} {9} (\bibinfo {year}
  {2013})}\BibitemShut {NoStop}%
\bibitem [{\citenamefont {Shidling}\ \emph {et~al.}(2019)\citenamefont
  {Shidling}, \citenamefont {Kolhinen}, \citenamefont {Schroeder},
  \citenamefont {Morgan}, \citenamefont {Ozmetin},\ and\ \citenamefont
  {Melconian}}]{Shid19}%
  \BibitemOpen
  \bibfield  {author} {\bibinfo {author} {\bibfnamefont {P.~D.}\ \bibnamefont
  {Shidling}}, \bibinfo {author} {\bibfnamefont {V.~S.}\ \bibnamefont
  {Kolhinen}}, \bibinfo {author} {\bibfnamefont {B.}~\bibnamefont {Schroeder}},
  \bibinfo {author} {\bibfnamefont {N.}~\bibnamefont {Morgan}}, \bibinfo
  {author} {\bibfnamefont {A.}~\bibnamefont {Ozmetin}},\ and\ \bibinfo {author}
  {\bibfnamefont {D.}~\bibnamefont {Melconian}},\ }\href@noop {} {\bibfield
  {journal} {\bibinfo  {journal} {Hyp. Int.}\ }\textbf {\bibinfo {volume}
  {240}} (\bibinfo {year} {2019})}\BibitemShut {NoStop}%
\bibitem [{\citenamefont {Ascher}\ \emph {et~al.}(2016)\citenamefont {Ascher},
  \citenamefont {Blank}, \citenamefont {Gerbaux}, \citenamefont {Giovinazzo},
  \citenamefont {Gr{\'e}vy}, \citenamefont {Kurtukian~Nieto}, \citenamefont
  {Severijns}, \citenamefont {Araujo-Escalona}, \citenamefont {Ban},
  \citenamefont {Fl{\'e}chard}, \citenamefont {Li{\'e}nard}, \citenamefont
  {Qu{\'e}m{\'e}ner},\ and\ \citenamefont {Zakoucky}}]{Asch16}%
  \BibitemOpen
  \bibfield  {author} {\bibinfo {author} {\bibfnamefont {P.}~\bibnamefont
  {Ascher}}, \bibinfo {author} {\bibfnamefont {B.}~\bibnamefont {Blank}},
  \bibinfo {author} {\bibfnamefont {M.}~\bibnamefont {Gerbaux}}, \bibinfo
  {author} {\bibfnamefont {J.}~\bibnamefont {Giovinazzo}}, \bibinfo {author}
  {\bibfnamefont {S.}~\bibnamefont {Gr{\'e}vy}}, \bibinfo {author} {\bibfnamefont
  {T.}~\bibnamefont {Kurtukian~Nieto}}, \bibinfo {author} {\bibfnamefont
  {N.}~\bibnamefont {Severijns}}, \bibinfo {author} {\bibfnamefont
  {V.}~\bibnamefont {Araujo-Escalona}}, \bibinfo {author} {\bibfnamefont
  {G.}~\bibnamefont {Ban}}, \bibinfo {author} {\bibfnamefont {X.}~\bibnamefont
  {Fl{\'e}chard}}, \bibinfo {author} {\bibfnamefont {E.}~\bibnamefont {Li{\'e}nard}},
  \bibinfo {author} {\bibfnamefont {G.}~\bibnamefont {Qu{\'e}m{\'e}ner}},\ and\
  \bibinfo {author} {\bibfnamefont {D.}~\bibnamefont {Zakoucky}},\ }\href
  {https://cds.cern.ch/record/2222309} {\emph {\bibinfo {title} {{WISARD:
  Weak-interaction studies with $^{32}$Ar decay}}}},\ \bibinfo {type} {Tech.
  Rep.}\ \bibinfo {number} {CERN-INTC-2016-050. INTC-I-172}\ (\bibinfo
  {institution} {CERN},\ \bibinfo {address} {Geneva},\ \bibinfo {year}
  {2016})\BibitemShut {NoStop}%
\bibitem [{\citenamefont {Araujo-Escalona}\ \emph {et~al.}(2019)\citenamefont
  {Araujo-Escalona}, \citenamefont {Atanasov}, \citenamefont {Fl{\'e}chard},
  \citenamefont {Alfaurt}, \citenamefont {Ascher}, \citenamefont {Blank},
  \citenamefont {Daudin}, \citenamefont {Gerbaux}, \citenamefont {Giovinazzo},
  \citenamefont {Gr{\'e}vy}, \citenamefont {Kurtukian-Nieto}, \citenamefont
  {Li{\'e}nard}, \citenamefont {Qu{\'e}m{\'e}ner}, \citenamefont {Severijns},
  \citenamefont {Vanlangendonck}, \citenamefont {Versteegen},\ and\
  \citenamefont {Zakoucky}}]{Arau20}%
  \BibitemOpen
  \bibfield  {author} {\bibinfo {author} {\bibfnamefont {V.}~\bibnamefont
  {Araujo-Escalona}}, \bibinfo {author} {\bibfnamefont {D.}~\bibnamefont
  {Atanasov}}, \bibinfo {author} {\bibfnamefont {X.}~\bibnamefont {Fl{\'e}chard}},
  \bibinfo {author} {\bibfnamefont {P.}~\bibnamefont {Alfaurt}}, \bibinfo
  {author} {\bibfnamefont {P.}~\bibnamefont {Ascher}}, \bibinfo {author}
  {\bibfnamefont {B.}~\bibnamefont {Blank}}, \bibinfo {author} {\bibfnamefont
  {L.}~\bibnamefont {Daudin}}, \bibinfo {author} {\bibfnamefont
  {M.}~\bibnamefont {Gerbaux}}, \bibinfo {author} {\bibfnamefont
  {J.}~\bibnamefont {Giovinazzo}}, \bibinfo {author} {\bibfnamefont
  {S.}~\bibnamefont {Gr{\'e}vy}}, \bibinfo {author} {\bibfnamefont
  {T.}~\bibnamefont {Kurtukian-Nieto}}, \bibinfo {author} {\bibfnamefont
  {E.}~\bibnamefont {Li{\'e}nard}}, \bibinfo {author} {\bibfnamefont
  {G.}~\bibnamefont {Qu{\'e}m{\'e}ner}}, \bibinfo {author} {\bibfnamefont
  {N.}~\bibnamefont {Severijns}}, \bibinfo {author} {\bibfnamefont
  {S.}~\bibnamefont {Vanlangendonck}}, \bibinfo {author} {\bibfnamefont
  {M.}~\bibnamefont {Versteegen}},\ and\ \bibinfo {author} {\bibfnamefont
  {D.}~\bibnamefont {Zakoucky}},\ }\href@noop {} {\bibinfo {title}
  {Simultaneous measurements of the beta neutrino angular correlation in
  $^{32}$ar pure fermi and pure gamow-teller transitions using beta-proton
  coincidences}} (\bibinfo {year} {2019}),\ \Eprint
  {https://arxiv.org/abs/1906.05135} {arXiv:1906.05135 [nucl-ex]} \BibitemShut
  {NoStop}%
\bibitem [{\citenamefont {Shrock}(1980)}]{Schr80}%
  \BibitemOpen
  \bibfield  {author} {\bibinfo {author} {\bibfnamefont {R.~E.}\ \bibnamefont
  {Shrock}},\ }\href@noop {} {\bibfield  {journal} {\bibinfo  {journal} {Phys.
  Let.}\ }\textbf {\bibinfo {volume} {96B}},\ \bibinfo {pages} {159} (\bibinfo
  {year} {1980})}\BibitemShut {NoStop}%
\bibitem [{\citenamefont {Bryman}\ and\ \citenamefont {Shrock}(2019)}]{Brym19}%
  \BibitemOpen
  \bibfield  {author} {\bibinfo {author} {\bibfnamefont {D.~A.}\ \bibnamefont
  {Bryman}}\ and\ \bibinfo {author} {\bibfnamefont {R.}~\bibnamefont
  {Shrock}},\ }\href@noop {} {\bibfield  {journal} {\bibinfo  {journal} {Phys.
  Rev. D}\ }\textbf {\bibinfo {volume} {100}},\ \bibinfo {pages} {073011}
  (\bibinfo {year} {2019})}\BibitemShut {NoStop}%
\end{thebibliography}

% here I copy/paste the bbl file made for me

\providecommand{\noopsort}[1]{}\providecommand{\singleletter}[1]{#1}%

\end{document}